\documentclass[bibyear,longauth]{aa}
\usepackage{graphicx}
\usepackage{txfonts}
\newcommand{\mincir}{\raise -2.truept\hbox{\rlap{\hbox{$\sim$}}\raise5.truept
\hbox{$<$}\ }}
\newcommand{\magcir}{\raise -2.truept\hbox{\rlap{\hbox{$\sim$}}\raise5.truept
\hbox{$>$}\ }}
\newcommand{\siml}{\raise -2.truept\hbox{\rlap{\hbox{$\sim$}}\raise5.truept
\hbox{$<$}\ }}
\newcommand{\simg}{\raise -2.truept\hbox{\rlap{\hbox{$\sim$}}\raise5.truept
\hbox{$>$}\ }}
\newcommand{\be}{\begin{equation}}
\newcommand{\ee}{\end{equation}}
\newcommand{\ba}{\begin{eqnarray}}
\newcommand{\ea}{\end{eqnarray}}

\newcommand {\h} {$h_{70}^{-1}$ Mpc$\;$}
\newcommand {\hh} {$h_{70}^{-1}$ Mpc}

\newcommand {\ks} {km~s$^{-1} \;$}
\newcommand {\kss} {km~s$^{-1}$}

\newcommand{\degree}{\ensuremath{\mathrm{^\circ}\;}}
\newcommand{\degreee}{\ensuremath{\mathrm{^\circ}}}
\newcommand{\arcm}{\ensuremath{\mathrm{^\prime}\;}}
\newcommand{\arcs}{\ensuremath{\arcmm\hskip -0.1em\arcmm \;}}
\newcommand{\arcmm}{\ensuremath{\mathrm{^\prime}}}

\newcommand{\dotsec}{\,\rlap{\hbox{$\mathrm{^s}$}}{\hbox{$.$}}\,}

\begin{document}

\title{CLASH-VLT: Substructure in the galaxy cluster MACS~J1206.2-0847
  from kinematics of galaxy populations \thanks{Based in large part on
    data acquired at the ESO VLT (prog.ID 186.A-0798).}}

\author{M. Girardi\inst{\ref{girar},\ref{bivia}}
\and A. Mercurio\inst{\ref{mercu}}
\and I. Balestra\inst{\ref{bivia}}
\and M. Nonino\inst{\ref{bivia}} 
\and A. Biviano\inst{\ref{bivia}} 
\and C. Grillo\inst{\ref{grill}}
\and P. Rosati\inst{\ref{rosat}}
\and M. Annunziatella\inst{\ref{girar},\ref{bivia}}
\and R. Demarco\inst{\ref{demar}}
\and A. Fritz\inst{\ref{scode}}
\and R. Gobat\inst{\ref{gobat}} 
\and D. Lemze\inst{\ref{lemze}} 
\and V. Presotto\inst{\ref{girar}}
\and M. Scodeggio\inst{\ref{scode}}
\and P. Tozzi\inst{\ref{tozzi}}
\and G. Bartosch Caminha\inst{\ref{rosat}}
\and M. Brescia\inst{\ref{mercu}}
\and D. Coe\inst{\ref{postm}}
\and D. Kelson\inst{\ref{kelso}}
\and A. Koekemoer\inst{\ref{postm}}
\and M. Lombardi\inst{\ref{lomba}}
\and E. Medezinski\inst{\ref{lemze}}
\and M. Postman\inst{\ref{postm}}
\and B. Sartoris\inst{\ref{girar},\ref{bivia}}
\and K. Umetsu\inst{\ref{umets}}
\and A. Zitrin\inst{\ref{zitri1},\ref{zitri2}}
\and W. Boschin\inst{\ref{boschi},\ref{boschi2},\ref{boschi3}}
\and O. Czoske\inst{\ref{ziegl}}
\and G. De Lucia\inst{\ref{bivia}} 
\and U. Kuchner\inst{\ref{ziegl}}
\and C. Maier\inst{\ref{ziegl}}
\and M. Meneghetti\inst{\ref{meneg1},\ref{meneg2},\ref{meneg3}}
\and P. Monaco\inst{\ref{girar},\ref{bivia}}
\and A. Monna\inst{\ref{seitz1}}
\and E. Munari\inst{\ref{girar},\ref{bivia}}
\and S. Seitz\inst{\ref{seitz1},\ref{seitz2}}
\and M. Verdugo\inst{\ref{ziegl}}
\and B. Ziegler\inst{\ref{ziegl}}
}

   \offprints{M. Girardi, \email{girardi@oats.inaf.it}}

\institute{Dipartimento di Fisica, Universit\`a degli Studi di Trieste, via Tiepolo 11, I-34143 Trieste, Italy\label{girar} 
\and INAF/Osservatorio Astronomico di Trieste, via G. B. Tiepolo 11, I-34133, Trieste, Italy\label{bivia} 
\and INAF/Osservatorio Astronomico di Capodimonte, Via Moiariello 16 I-80131 Napoli, Italy\label{mercu} 
\and Dark Cosmology Centre, Niels Bohr Institute, University of Copenhagen, Juliane Maries Vej 30, 2100 Copenhagen, Denmark\label{grill} 
\and Dipartimento di Fisica e Scienze della Terra, Universit\`a degli Studi di Ferrara, via Saragat 1, I-44122, Ferrara, Italy\label{rosat}
\and Department of Astronomy, Universidad de Concepcion, Casilla 160-C, Concepcion, Chile \label{demar} 
\and INAF/IASF-Milano, via Bassini 15, 20133 Milano, Italy\label{scode} 
\and Korea Institute for Advanced Study, KIAS, 85 Hoegiro, Dongdaemun-gu
Seoul 130-722, Republic of Korea\label{gobat} 
\and Department of Physics and Astronomy, The Johns Hopkins University, 3400 North Charles Street, Baltimore, MD 21218, USA\label{lemze} 
\and INAF/Osservatorio Astrofisico di Arcetri, Largo E. Fermi 5, 50125 Firenze, Italy\label{tozzi}
\and Space Telescope Science Institute, 3700 San Martin Drive, Baltimore, MD 21218, USA\label{postm} 
\and Observatories of the Carnegie Institution of Washington, Pasadena, CA 91 101, USA\label{kelso} 
\and Dipartimento di Fisica, Universit\`a degli Studi di Milano, via Celoria 16, I-20133 Milan, Italy\label{lomba} 
\and Institute of Astronomy and Astrophysics, Academia Sinica, P. O. Box 23-141, Taipei 10617, Taiwan\label{umets} 
\and California Institute of Technology, MC 249-17, Pasadena, CA 91125,
USA\label{zitri1}
\and Hubble Fellow\label{zitri2}
\and Fundaci\`on Galileo Galilei - INAF, Rambla Jos\'e Ana Fern\'andez P\'erez 7, E-38712 Bre\~na Baja, La Palma, Spain\label{boschi}
\and Instituto de Astrof\'isica de Canarias, C/V\'ia L\'actea s/n, E-38205 La Laguna, Tenerife, Spain\label{boschi2} 
\and Universidad de La Laguna, Dpto. Astrof\'isica, E-38206 La Laguna, Tenerife,
Spain\label{boschi3} 
\and University of Vienna, Department of Astrophysics, T\"urkenschanzstr. 17, 1180 Wien, Austria\label{ziegl} 
\and INAF/Osservatorio Astronomico di Bologna, via Ranzani 1, I-40127 Bologna, Italy\label{meneg1} 
\and INFN, Sezione di Bologna; Via Ranzani 1, I-40127 Bologna, Italy\label{meneg2}
\and Jet Propulsion Laboratory, California Institute of Technology, 4800
Oak Grove Drive, Pasadena, CA 91109\label{meneg3}
\and University Observatory Munich, Scheinerstrasse 1, D-81679 M\"unchen, Germany\label{seitz1} 
\and Max-Planck-Institut f\"ur extraterrestrische Physik, Posfatch 1312, Giessenbachstr., D-85741 Garching, Germany \label{seitz2}
}

\date{Received  / Accepted }

\abstract{} {In the effort to understand the link between the
  structure of galaxy clusters and their galaxy populations, we focus
  on MACSJ1206.2-0847 at $z \sim 0.44$ and probe its substructure in
  the projected phase space through the spectrophotometric properties
  of a large number of galaxies from the CLASH-VLT survey.}  {Our
  analysis is mainly based on an extensive spectroscopic dataset of
  445 member galaxies, mostly acquired with VIMOS@VLT as part of our
  ESO Large Programme, sampling the cluster out to a radius $\sim
  2R_{200}$ (4 \hh).  We classify 412 galaxies as passive, with strong
  H$\delta$ absorption (red and blue galaxies), and with emission
  lines from weak to very strong.  A number of tests for substructure
  detection are applied to analyze the galaxy distribution in the
  velocity space, in 2D space, and in 3D projected phase-space.}
         {Studied in its entirety, the cluster appears as a
           large-scale relaxed system with a few secondary, minor
           overdensities in 2D distribution.  We detect no velocity
           gradients or evidence of deviations in local mean
           velocities.  The main feature is the WNW-ESE elongation.
           The analysis of galaxy populations per spectral class
           highlights a more complex scenario.  The passive galaxies
           and red strong H$\delta$ galaxies trace the cluster center
           and the WNW-ESE elongated structure. The red strong
           H$\delta$ galaxies also mark a secondary, dense peak $\sim
           2$ \h at ESE.  The emission line galaxies cluster in
           several loose structures, mostly outside $R_{200}$.  Two of
           these structures are also detected through our 3D
           analysis. The observational scenario agrees with MACS
           J1206.2-0847 having WNW-ESE as the direction of the main
           cluster accretion, traced by passive galaxies and red
           strong H$\delta$ galaxies. The red strong H$\delta$
             galaxies, interpreted as poststarburst galaxies, date a
           likely important event 1-2 Gyr before the epoch of
           observation.  The emission line galaxies trace a secondary,
           ongoing infall where groups are accreted along several
           directions.}{}


\titlerunning{CLASH-VLT: Substructure in MACS~J1206.2-0847}

\maketitle
%

\section{Introduction}
\label{intro}

In the hierarchical scenario for large-scale structure formation,
clusters of galaxies are not relaxed structures.  Numerical
simulations show that clusters form preferentially through anisotropic
accretion of subclusters along the large-scale structure (LSS)
filaments (Colberg et al. \cite{col98}, \cite{col99} and
references therein).  From the observational point of view, it is well
known that a large fraction of clusters (30\%-70\%) contain
substructures, as shown by studies based on optical (e.g., Baier \&
Ziener \cite{bai77}; Geller \& Beers \cite{gel82}; Girardi et
al. \cite{gir97}; Ramella et al. \cite{ram07}; Wen \& Han \cite{wen13}), on X-ray data (e.g.,
Jones \& Forman \cite{jon99}; Buote \cite{buo02}; Jeltema et
  al. \cite{jel05}; Zhang et al. \cite{zha09}, B\"ohringer et
  al. \cite{boe10}), and on gravitational lensing techniques (e.g.,
Athreya et al. \cite{ath02}; Dahle et al. \cite{dah02}; Smith et
al. \cite{smi05}: Grillo et al. \cite{gri14a}), indicating that past
signatures of cluster accretion are quite common.  Much progress has
been made in the study of cluster accretion phenomena (see Feretti et
al. \cite{fer02} for a general review) and recent, dedicated studies
have often focused on a few major, ongoing cluster mergers, for
instance in the context of Dynamical Analysis Radio Clusters project
(DARC, e.g., Girardi et al. \cite{gir11}) and MUlti-Wavelength Sample
of Interacting Clusters project (MUSIC, Maurogordato et
al. \cite{mau11}). Other dedicated studies have focused on larger
scales, i.e.,  on cluster accretion through filaments (e.g., Fadda et
al.  \cite{fad08}; Dietrich et al. \cite{die12}).  On-going cluster
formation is also evident in distant clusters as pioneering studies
have shown that most clusters identified at $z\geq 0.8$ are elongated,
clumpy, or with a filamentary structure (e.g., Donahue et
al. \cite{don98}; Gioia et al. \cite{gio99}; Demarco et
al. \cite{dem07}; Fassbender et al. \cite{fas11}).

It is well established that the properties of cluster galaxies differ
from those of field galaxies and that clusters are characterized by
radial gradients. Galaxies in denser, central regions are
usually of earlier morphological type, redder color,  and lower
star formation rate (hereafter SFR; e.g., Gerken et
al. \cite{ger04}). However, the precise details of the connection
between galaxy evolution and cluster environment are still
unknown. Several physical mechanisms can be involved in modifying
galaxies in the cluster environment (see Fig.~10 by Treu et
al. \cite{tre03} and refs. therein). Moreover, it is clear that a
scenario in which galaxies are accreted in a static cluster
environment is too simplistic.  In fact, there  is evidence of a
  connection between the fraction of active galaxies and the presence
  and/or the amount of substructure, generally interpreted as an enhanced star formation (SF)  in member galaxies caused by the process of cluster mergers (e.g.,
Caldwell \& Rose 1997; Ferrari et al. \cite{fer05}; Owen et
al. \cite{owe05}). An alternative interpretation is that
  quenching SF is less effective in just forming multicomponent
  clusters (see Cohen et al. \cite{coh14} for further
  discussions). From a theoretical point of view, during a major
cluster merger, the SF in gas-rich galaxies can decrease or,
alternatively, be triggered depending on the specific interaction
between the intracluster medium (ICM) and the interstellar gas within
the galaxies (e.g., Fujita et al. \cite{fuj99}; Bekki et al.
\cite{bek10}).  Time-dependent tidal fields of merging groups and
clusters of galaxies can also trigger secondary starbursts in their
member galaxies (Bekki \cite{bek99}).

In this context, poststarburst galaxies (PSBs, Dressler \& Gunn
\cite{dre83}) play a crucial role. In fact, their spectral signature,
i.e., strong Balmer absorption lines and weak or no emission lines, is
generally attributed to a vigorous SF that has recently decreased
significantly or ceased altogether.  More properly, one can think of a
real poststarburst phase or, alternatively, a star-forming truncated
phase (e.g., Dressler et al. \cite{dre09}).  The last SF event in PSBs
galaxies ended  a few Myr to 2 Gyr before the time of
observation, on the basis of the lifetime of the stars responsible for
the strong Balmer lines (e.g., Poggianti \& Barbaro \cite{pog96},
\cite{pog97}; Balogh et al. \cite{bal99}; Poggianti et
al. \cite{pog99}; Mercurio et al. \cite{mer04}, hereafter M04).  More
specifically, the typical time elapsed since the last SF in the red
PSBs lies in the range 1-2 Gyr, while in the blue PSBs the
interruption of SF occurred within the last $0.5$ Gyr (M04).  The
location of PSBs in the parent cluster can give important information
on the driving physical mechanism.  For instance, the synchronized
triggered SF in the cluster merging simulations of Bekki et al.
(\cite{bek10}) gives origin to a population of PSBs having a spatial
distribution different from the rest of the galaxy population.
Analyzing nine clusters at $z\sim 1$ drawn from the GCLASS survey,
Muzzin et al. (\cite{muz14}) have found that PSBs appear to trace a
coherent ring in the projected phase space.  Alternatively, Dressler
et al. (\cite{dre13}) have found that in the IMACS Cluster Building Survey
(five rich clusters at $0.31<z<0.55$) the PSB population traces the same
spatial distribution of the passive population.

Recently, significant interest has  focused on cluster outskirts, too.
In fact, although there is evidence that SF is suppressed on average
 outside 2$R_{200}$\footnote{The radius $R_{\delta}$ is the radius
  of a sphere with mass overdensity $\delta$ times the critical
  density at the redshift of the galaxy system.} in both  low- and high-redshift clusters (e.g., Lewis et al. \cite{lew02}; Gomez et
al. \cite{gom03}; Rasmussen et al. \cite{ras12}), several studies
have associated starburst features with infalling groups (e.g., Owen et
al. \cite{owe05}; M04; Poggianti et al. \cite{pog04}; Oemler et
al. \cite{oem09}). In  nearby clusters an increase of SF along the filaments feeding the clusters has been found
(e.g., Fadda et al. \cite{fad08}). This seems particularly true 
around 1.5-2$R_{200}$ and for galaxies belonging to groups, suggesting
that a relatively high density in the infalling regions promotes
interactions between galaxies (e.g., galaxy-galaxy harassment) and
following momentary bursts of SF (Scott et al. \cite{por07}; Mahajan et
al. \cite{mah12}).  Recent simulations suggest that cluster galaxies
may be pre-processed before the infall into the cluster environment,
although the postprocessing is important as well (Vijayaraghavan \&
Ricker \cite{vin13}).

Optical multiband  data and particularly multiobject spectroscopy are
a consolidated way to investigate cluster substructure and cluster
merging phenomena (e.g., Girardi \& Biviano \cite{gir02}).  The few
recent studies of clusters with hundreds of spectroscopic members have
shown the power of using large spectroscopic catalogs to study cluster
internal structure (e.g., Owers et al. \cite{owe11}; Munari et
al. \cite{mun14}). In some studies, the correlation of passive and
active galaxies populations with the substructure of the parent
cluster has been investigated (e.g., Czoske et al. \cite{czo02}; M04;
Mercurio et al.  \cite{mer08}; Oemler et al. \cite{oem09}; Ma et
al. \cite{ma10}). Very large spectroscopic datasets are needed to
study cluster substructure in phase space for different spectral types
and to infer cluster assembly history. The present study represents a
pilot study in this new direction. It focuses on the galaxy cluster
\object{MACS J1206.2-0847} (hereafter MACS1206; Ebeling et
al. \cite{ebe09}) at $z \sim 0.44$, discovered and first
  described in the REFLEX Galaxy Cluster Survey catalog (\object{RXC
    J1206.2-0848}; B\"ohringer et al. \cite{boe04}).

\subsection{MACS1206 and the CLASH-VLT programme}
\label{introclash}

The subject of this study, the galaxy cluster MACS1206 is part of the
survey of 25 target clusters of the HST Multi-Cycle Treasury program
``Cluster Lensing And Supernova'' (CLASH, P.I.: M. Postman; Postman et
al. \cite{pos12}). Our study is based on an extensive spectroscopic
dataset collected within the ESO Large Programme 186.A-0798 ``Dark
Matter Mass Distributions of Hubble Treasury Clusters and the
Foundations of $\Lambda$CDM Structure Formation Models'' (hereafter
CLASH-VLT, P.I.: P. Rosati; Rosati et al. \cite{ros14}). This program
aims at obtaining spectra for at least 500 cluster members for each of
the 14 southern CLASH clusters with $0.2 < z < 0.6$, using the VIsible
Multi-Object Spectrograph (VIMOS) at VLT (CLASH-VLT survey). The  study
of the combined evolution of galaxies and parent cluster requires i)
the spectroscopic identification of a large number of cluster galaxies
in each cluster; ii) sampling a region out to at least twice
$R_{200}$; and  iii) a range of cluster redshifts covering a relevant epoch
cluster formation. Specifically, the past 5 Gyr play a crucial role in
cluster assembly history since the overall morphological content, the
fraction of galaxies of different spectral types, and the SF history
of cluster galaxies are known to significantly evolve since $z\sim0.5$
(e.g., Butcher \& Oemler \cite{but84}; Couch \& Sharples \cite{cou87};
Dressler et al. \cite{dre97}; Fritz et al. \cite{fri05}; Poggianti et
al. \cite{pog09}). Useful complementary information from
optical, X-ray and radio data, and high-resolution HST imaging is
available, too. Other CLASH-based studies, focused on other aspects of
MACS1206, have already been presented and are referred to and discussed
throughout this paper (Umetsu et al.~\cite{ume12}; Zitrin et
al.~\cite{zit12}; Biviano et al.~\cite{biv13}, B13; Annunziatella et
al.~\cite{ann14}; Grillo et al.~\cite{gri14b}; Presotto et
al.~\cite{pre14}).

The paper is organized as follows. We present our catalog and galaxy
spectral type classification in Sects.~2 and 3, respectively.  Section~4
concentrates on the analysis of cluster structure.  Section~5 is devoted
to the interpretation and discussion of our results.  We give our
summary and conclusions in Sect.~6.

Unless otherwise stated, we give errors at the 68\% confidence level
(hereafter c.l.). Throughout this paper, we use $H_0=70$ km s$^{-1}$
Mpc$^{-1}$ in a flat cosmology with $\Omega_0=0.3$ and
$\Omega_{\Lambda}=0.7$. In the adopted cosmology, 1\arcm corresponds
to $0.341$ \h at the cluster redshift.

\section{Redshift catalog and member selection}
\label{cat}

The full redshift catalog of MACS1206 consists of 3292 objects, with
measured redshifts mostly acquired as part of our ESO Large Programme
186.A-0798 (P.I.: Piero Rosati) using the VIMOS@VLT/UT3.  Additional
archival VIMOS data have been homogeneously reduced from programs
169.A-0595 (P.I.: Hans B\"ohringer) and 082.A-0922 (P.I. Mike
Lerchster) and have provided 952 spectra in the cluster field. All
data are reduced with the same VIPGI package (see Scodeggio et
al. \cite{sco05}).  Four redshift quality classes were defined
``secure'' (QF=3), ``based on a single-emission-line'' (QF=9), ``likely''
(QF=2), and ``insecure'' (QF=1), having $> 99.99\%$,
$\sim 92\%$, $\sim 75\%$, and $< 40\%$ probability to be correct, respectively.
Further details on the ESO Large Programme 186.A-0798 are reported
elsewhere [Rosati et al. (\cite{ros14}) and in Rosati et al. (in
prep.)]. The catalog also includes 22 spectra acquired with FORS2@VLT
as part of ESO Programme ID 089.A-0879 (P.I.: Raphael Gobat) and
additional literature redshifts taken from Lamareille et
al. (\cite{lam06}, 1 object), Jones et al. (\cite{jon04}, 3 objects),
Ebeling et al. (\cite{ebe09}, 25 objects), and 21 objects observed with
IMACS-GISMO at the Magellan telescope (Daniel Kelson private
communication).  The full catalog, with the exception of QF=1
redshifts has already been used by B13 and Lemze et
al.~(\cite{lem13}) and is publicly available.
In this paper, we do not consider redshifts with QF=1 and 2.  Our
reference sample includes 1920 galaxies.  

Biviano et al.~(\cite{biv13}) quantified the completeness of the
spectroscopic catalog by computing the ratio between the $R_{\rm
  C}$-band number counts of objects with measured redshift and all
photometric objects.  In particular, B13 show that in the virial
region the completeness is $\sim 0.6$ at $R_{\rm C}\sim 20$, $\sim
0.3$ at $R_{\rm C}\sim 23$, and then declines very sharply (see their
Fig.~4).  However, although the redshift catalog covers a fraction of
the photometric objects, it is essentially unbiased with respect to
galaxy color. Biviano et al.~(\cite{biv13}) also evaluated the
spatial completeness of the spectroscopic sample finding a good
uniformity with a mild radially-dependent incompleteness. The
completeness varies with radius from 0.6 at the cluster center to 0.2
at the external cluster limit (see their Fig.~5).  Only two
low-overdensity features are shown in their Fig.~5: the first in the
center, elongated towards the south direction, the second
10\arcmin\ to the east.  Our sample and that of B13 do not differ in
their spatial distributions according to the 2D Kolmogorov-Smirnov
test (hereafter 2DKS-test, Fasano \& Franceschini \cite{fas87}). A
posteriori, we verified that none of the galaxy clumps or subclusters
we discuss in this paper resemble the specific overdensity features
detected in the completeness map by B13 (cf. their Fig.~5 with our
results and figures in Sect.~3.2).  We conclude that our results are
not affected by the redshift spatial incompleteness.

To select cluster members, we applied the two-step method
called ``peak+gap'' (P+G) already applied by B13 and previous studies (e.g.,
Girardi et al. \cite{gir11} and refs. therein). The method is a
combination of the 1D adaptive-kernel method DEDICA (Pisani
\cite{pis93}) and the ``shifting gapper'', which uses both position and
velocity information (see Fadda et al. \cite{fad96}; Girardi et
al. \cite{gir96}).  For the center of MACS1206, we adopted the
position of the brightest cluster galaxy [BCG,
  R.A.=$12^{\mathrm{h}}06^{\mathrm{m}}12\dotsec15$,
  Dec.=$-08\degreee48\arcmm 03.3\arcs$ (J2000)].  In the first step,
we detected MACS1206 as a peak at $z\sim0.4385$ populated by 466
galaxies. The second step leads to 445 cluster members.

By applying the biweight estimator to the 445 cluster members (Beers
et al. \cite{bee90}, ROSTAT software), we computed a mean cluster
line-of-sight (LOS) velocity $\left<V\right>=\left<cz\right>
=(131\,843\pm$49) \kss, corresponding to a mean cluster redshift
$\left<z\right>=0.4398\pm0.0002$.  We estimated the LOS velocity
dispersion, $\sigma_V$, by using the biweight estimator and applying
the cosmological correction and the standard correction for velocity
errors (Danese et al. \cite{dan80}).  We obtained
$\sigma_V=1035_{-45}^{+27}$ \kss, where errors are estimated through a
bootstrap technique. Both $\left<V\right>$ and $\sigma_V$ are in
agreement with the values obtained by B13 using a larger galaxy sample.
Out of the reference sample of 445 cluster members (hereafter the TOT
sample), we also considered the sample of 249 galaxies within
$R_{200}$ (hereafter the R200 sample). The
estimate of $R_{200}$ spans the range 1.96-2.08 \hh, with a typical
1$\sigma$ uncertainty of $\sim 5\%$, depending on the method of
analysis (see Table~2 of B13).  Hereafter the value $R_{200}=2$ \h is
assumed.

Throughout the paper we also use color and morphological information
coming from the Subaru Suprime-Cam data. These data were retrieved from
the SMOKA archive\footnote{http://smoka.nao.ac.jp/SUPsearch} and
reduced following the same steps as adopted by Nonino et
al. (\cite{non09}).  Further information can be found in Presotto et
al. (\cite{pre14}) and in Mercurio et al. (in prep.), where this
photometric information is more extensively used.

\section{Spectral type classification and galaxy populations}
\label{classpop}

\subsection{Spectral type classification}
\label{class}

Out of a total of 445 member galaxies, we have the spectra of 415
galaxies acquired at VIMOS@VLT (405 objects) and FORS2@VLT (10 objects),
other redshifts coming from the literature or other sources (see
Sect.~\ref{cat}).  Following most of the literature, we used
emission lines and H$\delta$ absorption lines to classify galaxies.
We used wavelength--calibrated and flux--calibrated galaxy spectra as
obtained from VIPGI (Scodeggio et al. \cite{sco05}).  We corrected for
galactic extinction following Schlegel et al. (\cite{sch98}) and
using the IRAF~\footnote{IRAF is distributed by the National Optical
  Astronomy Observatories, which are operated by the Association of
  Universities for Research in Astronomy, Inc., under cooperative
  agreement with the National Science Foundation.}  task DEREDDEN
included in the package ONEDSPEC. Finally, the calibrated and
dereddened spectra were corrected for the measured velocity dispersion
by using DISPCOR.

\begin{figure}
\centering
\resizebox{\hsize}{!}{\includegraphics{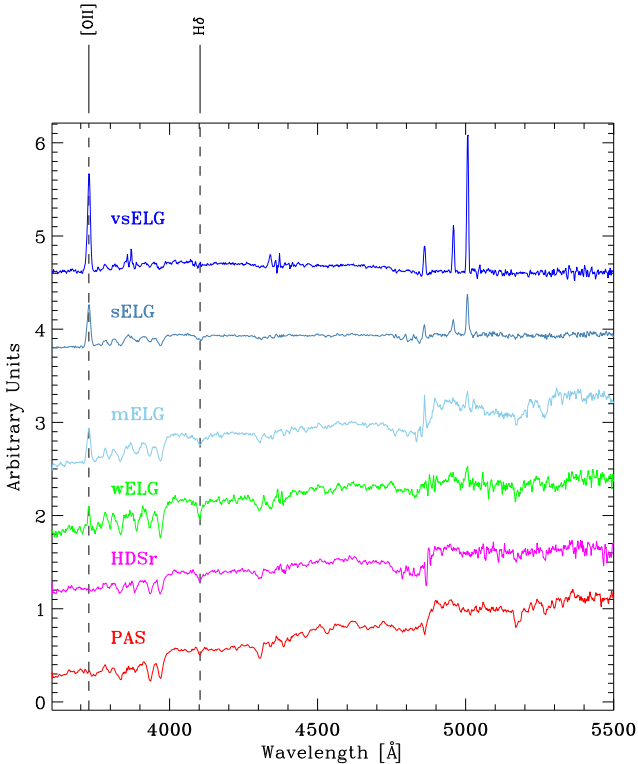}}
\caption
{Co-added (rest-frame) spectra of cluster galaxies per spectral
  class. The positions of [OII] and H$\delta$ lines used in the
  classification are indicated.
}
\label{figstackspectra}
\end{figure}

We measured the equivalent widths (EWs) for the emission lines [OII],
[OIII], and when available H${\alpha}$. All the spectra cover
  the [OII] region with the exception of 18 galaxies. We also
  measured the EW for H${\delta}$ (see M04 for the definition of
wavelength ranges).  We detected 185 galaxies with evidence of
emission lines. The 185 galaxies with emission lines were divided in
four classes depending on the strength of the [OII] emission line,
or of the [OIII] emission line for the four cases where the [OII]
  region is not covered. In these four cases we assumed a [OIII]/[OII]
  flux ratio equal to one.  We considered the wELG, mELG, sELG, and
vsELG classes, i.e., those formed of galaxies with weak, medium,
strong, and very strong emission lines (see Table~\ref{tabclass}).
Figure~\ref{figstackspectra} shows the co-added spectra for emission
line galaxies and other classes.  All the emission line galaxies, with
the exception of wELGs, show similar distributions of the
clustercentric distances (see Sect.~\ref{pop}), and we aggregated them
in the ELG class.

\begin{table*}
        \caption[]{Spectral type classification.}
         \label{tabclass}
                $$
         \begin{array}{l l r c c}
            \hline
            \noalign{\smallskip}
            \hline
            \noalign{\smallskip}
\mathrm{Galaxy\ characteristics} & \mathrm{Class} &\mathrm{N_g} & \mathrm{EW([OII])}^{\mathrm{a}}& \mathrm{EW(H{\delta})} \\
& & & &\\
            \hline
            \noalign{\smallskip}
\mathrm{Passive}                 &\mathrm{PAS}                   &183&\mathrm{absent}&\mathrm{absent}\\
\mathrm{Strong\ H{\delta}\ absorption\ \&\ red\ color}&      \mathrm{HDSr }^{\mathrm{b}}                 & 38&\mathrm{absent}&>3\mathrm{\AA}\\
\mathrm{Strong\ H{\delta}\ absorption\ \&\  blue\ color} &\mathrm{HDSb }^{\mathrm{b}}                 &  5&\mathrm{absent}&>5\mathrm{\AA}\\
\mathrm{Weak\ emission\ lines}                 & \mathrm{wELG }                 & 17&>-7\AA&\mathrm{any}\\
\mathrm{Medium\ emission\ lines}&\mathrm{mELG }^{\mathrm{c}}     & 24&(-15\mathrm{\AA},-7\mathrm{\AA}]&\mathrm{any}\\
\mathrm{Strong\ emission\ lines}&\mathrm{sELG }^{\mathrm{c}}     &101&(-40\mathrm{\AA},-15\mathrm{\AA}]&\mathrm{any}\\
\mathrm{Very\ strong\ emission\ lines}&\mathrm{vsELG}^{\mathrm{c}}     & 43&\le-40\mathrm{\AA}&\mathrm{any}\\
              \noalign{\smallskip}
            \hline
         \end{array}
$$ \tablefoot{(a)~In four cases EW([OIII]) were used, see 
             the text and  Table~\ref{tabcatesempio}; (b)~The HDS
           class ($N_g=44$) is formed of HDSr and HDSb galaxies, and
           an additional galaxy with strong H$\delta$ absorption,
           which can be classified as  neither HDSr nor HDSb; (c)~The ELG
           class ($N_g=168$) is formed of mELG, sELG, and vsELG.
           Throughout the paper we use PASs, HDSrs, HDSbs, and
           w/m/s/vsELGs to refer to the respective member galaxies.}
\end{table*}

Among the non-emission line galaxies, the 44 galaxies with
EW(H${\delta})>$3\AA\ were classified as strong H${\delta}$ absorption
galaxies (HDS sample), interpreted as PSBs or galaxies with truncated
SF (see Sect.~\ref{intro}).  Compared to  red galaxies, blue
galaxies require larger EW(H$\delta$) to be identified as galaxies
that have a peculiar star-forming activity.  Thus, a more precise way
to classify strong H$\delta$ galaxies uses the diagram of
EW(H$\delta$) vs ($B-R$) color or, alternatively, vs the
strength of the 4000 \AA\ break, which correlates with color (Couch \&
Sharples \cite{cou87}; Barger et al. \cite{bar96}; Balogh et
al. \cite{bal99}).  We considered 38 strong H${\delta}$ galaxies
with red colors and EW(H$\delta$)$>3$\AA\ (HDSr) and 5 strong
H${\delta}$ galaxies with blue colors and
EW(H$\delta$)$>5$\AA\ (HDSb). In this separation, we followed M04 using
their threshold in the differential color $(B-R_{\rm C})_{\rm
  diff}=-0.5$, where the differential color is defined as the observed
color minus the color of the color-magnitude relation at the given
magnitude.  The color-magnitude relation was obtained as fitted to the
spectroscopically confirmed members ($B-R_{\rm C}=3.587-0.0714\,R_{\rm
  C}$), where $R_{\rm C}$ indicates the Kron magnitudes and the
$B-R_{\rm C}$ color is computed on aperture magnitudes (Mercurio et
al. in prep.).  In their original study, M04 based their HDS
classification scheme directly on the strength of the 4000 \AA\ break,
while we use the corresponding color. The color-magnitude
correction allows us to take into account the large magnitude range
spanned by our dataset.

With  the exception of three galaxies whose  signal-to-noise
is too low according to our standards (S/N$\sim$2 for the H$\delta$ absorption
line), the remaining 183 galaxies were classified as passive (PAS).
We  note that the BCG is classified as a passive galaxy in spite of the
presence of a [OII] emission line in the VIMOS spectrum since the HST
photometric data suggest that this line can be associated with  a blue
compact source and/or peculiar features blended with the BCG (see
Presotto et al. \cite{pre14} for the relevant discussion).

Summarizing, we have spectral type classifications for 412 galaxies,
217 galaxies within $R_{200}$. The spectroscopic catalog is
electronically published in Table~\ref{tabcatesempio}, available at CDS.
Table~\ref{tabclass} summarizes the numbers of galaxies assigned to
each spectral class.

\begin{table*}[ht]
    \caption[]{Spectral classification of galaxies in the cluster MACS~J1206.2-0847}
\label{tabcatesempio}
           $$ 
           \begin{array}{r|c|c|c|c|c|c|c|}
            \hline
            \noalign{\smallskip}
            \hline
            \noalign{\smallskip}

\mathrm{ID} & \alpha(\mathrm{J2000})&\delta(\mathrm{J2000})& z & 
B-R_{\mathrm{C}} & \mathrm{EW([OII])} & \mathrm{EW(H{\delta})} & \mathrm{Class}\\
   &                       &                      &   &                 & \mathrm{\AA}  &\mathrm{\AA}      &      \\

\hline
\noalign{\smallskip}

 13889 & 12:06:55.15 & -08:56:29.7 & 0.4367 &   1.77\pm0.01 &   -1.50\pm   0.30 &    6.90\pm   0.20 & \mathrm{wELG} \\
 19586 & 12:06:05.16 & -08:53:20.6 & 0.4360 &   1.26\pm0.01 &   -2.80\pm   0.50 &    2.50\pm   0.50 & \mathrm{wELG} \\
\noalign{\smallskip}
\hline
\noalign{\smallskip}
\hline
           \end{array}
$$ \tablefoot{This table will be made available at the CDS. A portion
             is shown here for guidance regarding its form and
             content. Column~1: running ID for galaxies in the presented
             sample; Cols.~2 and 3: R.A. and Dec. (J2000); Col.~4:
             spectroscopic redshift, $z$; Col.~5 $B-R_{\rm C}$ color;
             Cols.~6-7: [OII] and H$\delta$ equivalent widths. We list
             `....' when the EW cannot be measured because the line does
             not lie within the available wavelength range.  We list `0.0' for EW[OII] when no emission is
               observed. Column~8: the spectral classification.}
\end{table*}

To compare our spectral-classification statistics to that of Dressler
et al.  (\cite{dre13}), which considers five rich clusters at
$0.31<z<0.54$, we also computed the fractions per spectral class
within 1.5 $R_{200}$ and $R_{\rm C}<22.3$.  We obtained the fraction
(PAS+HDS)/TOT=$67\%$ and HDS/PAS=$24\%$. These values do not change
significantly when taking into account the radial spatial
incompleteness (see Fig.~5 of B13 and our Sect.~\ref{cat}), i.e., normalizing the number of galaxies in different cluster radial
  bins ($R\le 0.4$ \hh, $0.4<R\le1$ \hh, $1<R\le 2$ \hh, $R >2$ \hh)
  using different spatial completeness corrections (i.e., dividing by
  0.6, 0.5, 0.4, 0.3).  Our values are in good agreement with the
corresponding fractions (PAS+PSB)/TOT and PSB/PAS reported by Dressler
et al.  (\cite{dre13}, see the cluster values in their Fig.~16 and
Table~4).

\begin{figure}
\centering
\resizebox{\hsize}{!}{\includegraphics{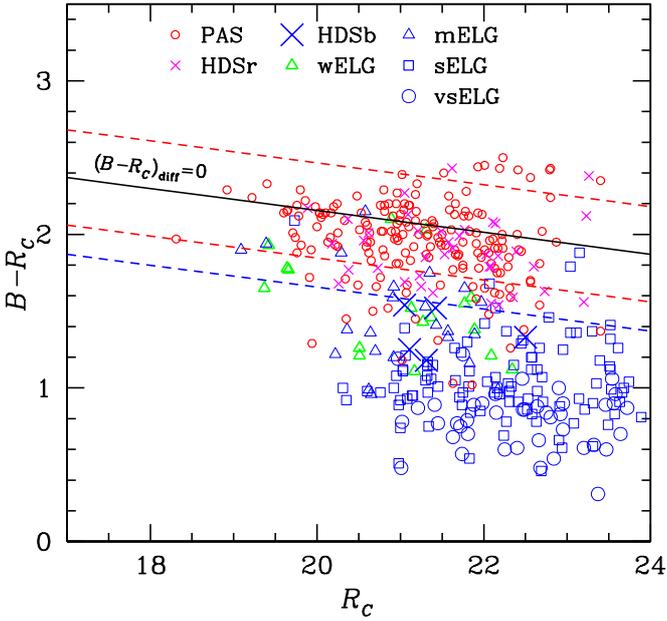}}
\caption { $B-R_{\rm C}$ vs $R_{\rm C}$ color-magnitude diagram with
  our seven spectral classes.  The black solid line is the
  color-magnitude relation, $(B-R_{\rm C})_{\rm diff}=0$. The blue
  dashed line indicates the ($B-R_{\rm C}$)$_{\rm diff}=-0.5$ value
  which we use to separate HDSr and HDSb following M04.  The two red
  dashed lines indicate the locus of the red sequence galaxies,
  $|(B-R_{\rm C})_{\rm diff}|<0.3$.}
\label{figcm}
\end{figure}

\begin{figure}
\centering
\resizebox{\hsize}{!}{\includegraphics{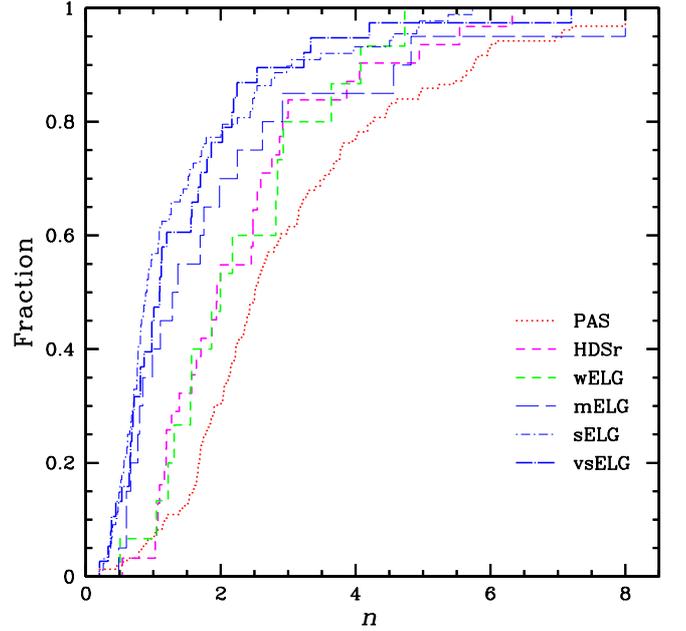}}
\caption
{Cumulative distribution of S\'ersic index of galaxies per spectral class.}
\label{figmatkwosersic}
\end{figure}

We also used some results from the morphological analysis of our
galaxy sample as derived from Suprime-Cam@Subaru data (Mercurio et
al., in prep.).  The PAS, HDSr, wELG, and ELG sequences of spectral
type populations correspond to bluer and bluer galaxies (see
Fig.~\ref{figcm}).  For each spectral class with the exception of 
HDSb, Figure~\ref{figmatkwosersic} shows the distribution of
S\'ersic index $n$, as determined in the $R_C$–band Subaru image using
the GALAPAGOS software (Barden et al. \cite{bar12}) and a single
S\'ersic profile.  The value of $n$ is larger for more prominent
bulges and $n$=2-2.5 can be considered the transition value between
disk-dominated and bulge-dominated galaxies (e.g., M04; Barden et al.
\cite{bar05}; Fisher \& Drory \cite{fis08}).  We note a good
correlation between the S\'ersic index $n$ and spectral classes, with
PASs mostly bulge-dominated and ELGs mostly disk-dominated, while HDSr
and wELG galaxies have intermediate values of $n$.

\subsection{Space and velocity distributions of galaxy populations}
\label{pop}

\begin{figure}
\centering 
\resizebox{\hsize}{!}{\includegraphics{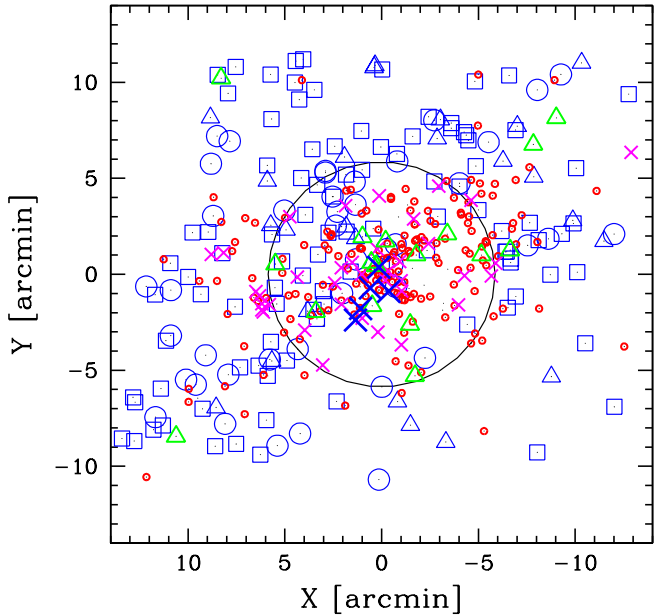}}
\caption
{Spatial distribution of the the 445 cluster members highlighting the
  spatial segregation between early- and late-type
  galaxies across our spectral classes. Each classified
  galaxy is marked by a symbol: PAS (small red circles); HDSr (magenta
  crosses); HDSb (large blue crosses); wELG (green triangles); mELG
  (blue triangles); sELG (blue squares); vsELG (blue circles).  The
  circle centered on the BCG encloses the $R_{200}$
  region.
}
\label{fig2dspec}
\end{figure}

\begin{figure}
\centering
\resizebox{\hsize}{!}{\includegraphics{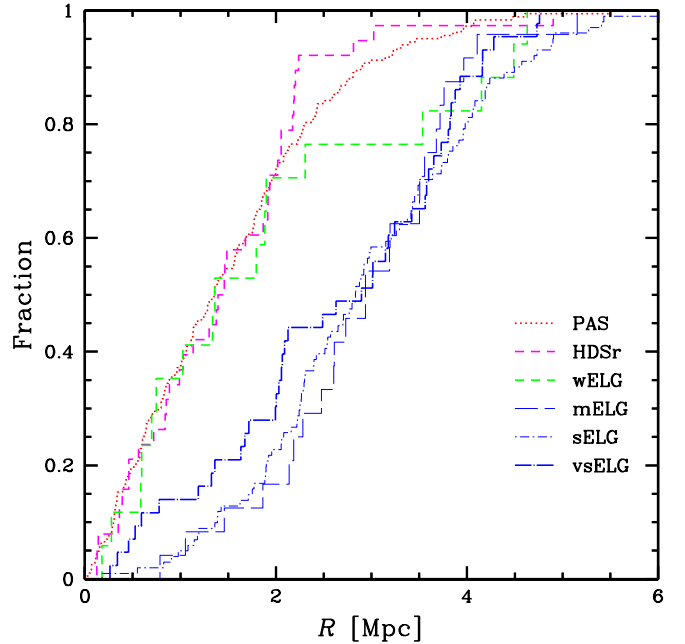}}
\caption
{Cumulative distributions of clustercentric distance $R$ of galaxies
  per spectral class, showing the spatial segregation from 
    early- through late-type  galaxies.  
}
\label{figmatkwor}
\end{figure}

Figure~\ref{fig2dspec} shows that PAS and HDSr galaxies are more
spatially clustered than  ELGs.  To make a quantitative
comparison among different galaxy populations, we applied the
Kruskall-Wallis test to the clustercentric distances $R$ (KW-test,
e.g., Ledermann \cite{led82}). This test is a non-parametric method
for testing whether samples originate from the same distribution. The
KW test leads to significant results ($>99.99\%$ c.l.); at least
one of the samples is different from the other samples.  In
Fig.~\ref{figmatkwor} there is a clear dichotomy between the
population formed of PAS, HDSr, and wELG galaxies and the population
formed of mELG, sELG, and vsELG galaxies, the galaxies of the former group being more
clustered.  We also compared spectral classes two by two. We applied
the 1D Kolmogorov-Smirnov test (hereafter 1DKS-test; e.g., Ledermann
\cite{led82}) to compare the distributions of clustercentric
distances.  When comparing one of the PAS, HDSr, wELG classes with one
of the mELG, sELG, vsELG classes we detected significant differences
(at $>98\%$ c.l.).  The comparison between PAS and HDSr, PAS and wELG,
and HDSr and wELG shows no differences, and similarly the comparison
between mELG and sELG, mELG and vsELG, and sELG and vsELG. The results of
the 1DKS-test confirm the dichotomy between the population formed of
PAS, HDSr, and wELG galaxies and that formed of mELG, sELG and vsELG
galaxies and prompted us to treat mELGs, sELGs, and vsELGs together in
the combined ELG class, separately from wELGs. This is a more detailed
view, across several spectral types, of the well-known spatial
segregation of spiral/elliptical galaxies and of blue/red galaxies in
local and distant clusters (e.g., Melnick \& Sargent \cite{mel77};
Dressler \cite{dre80}; Whitmore et al. \cite{whi93}; Abraham et
al. \cite{abr96}; Dressler et al. \cite{dre99}; Gerken et
al. \cite{ger04}).

In a second-order analysis, we applied the 2DKS-test to compare the
distributions of projected positions.  This is particularly meaningful
for the MACS1206 dataset, which is large enough to search for
differences in the internal structure.  This test is more sensitive
than the above 1DKS-test and, in addition to confirming the above dichotomy,
it is able to detect differences within the ELG class, specifically
between mELG and vsELG, and between sELG and vsELG (marginally, at the
$<95\%$ c.l.). The combined mELG$+$sELG class differs from vsELG at
the $96\%$ c.l.. The observed difference is related to the cluster
substructure and will be further analyzed and discussed in following
sections.

\begin{table}
        \caption[]{Kinematical  properties of the whole cluster and galaxy populations.}
         \label{tabv}
                $$
         \begin{array}{l r l l}
            \hline
            \noalign{\smallskip}
            \hline
            \noalign{\smallskip}
\mathrm{Sample} & \mathrm{N_g} & \phantom{249}\mathrm{<V>}\phantom{249} & 
\phantom{24}\sigma_V\phantom{24}\\
& &\phantom{249}\mathrm{km\ s^{-1}}\phantom{249}&\phantom{2}\mathrm{km\ s^{-1}}\phantom{24}\\
            \hline
            \noalign{\smallskip}
\mathrm{TOT}  &445&131843\pm49 &1035_{-45}^{+27}\\
\mathrm{R200 }  &249&131776\pm74 &1169_{-52}^{+54}\\
\mathrm{PAS  }      &183&131980\pm69 &\phantom{1}940_{-85}^{+76}\\
\mathrm{HDSr  }     & 38&131630\pm196&1188_{-379}^{+249}\\
\mathrm{HDSb  }     &  5&132836\pm767&1307_{-141}^{+979}\\
\mathrm{wELG }      & 17&131502\pm307&1215_{-363}^{+415}\\
\mathrm{ELG}^{\mathrm{a}}&168&131690\pm 78&1014_{-83}^{+83}\\
\mathrm{mELG }      & 24&132121\pm192&\phantom{1}917_{-299}^{+366}\\
\mathrm{sELG }      &101&131698\pm 99&\phantom{1}996_{-105}^{+90}\\
\mathrm{vsELG}      & 43&131485\pm170&1102_{-161}^{+235}\\
              \noalign{\smallskip}
            \hline
         \end{array}
$$ \tablefoot{Column~1: sample ID [(a)~The ELG class is formed of mELG, sELG, and
vsELG]; Col.~2: the number of assigned galaxies, $N_{\rm g}$; Cols.~3 and 4:
mean LOS velocity and velocity dispersion of galaxies, $\left<V\right>$ and $\sigma_V$, with respective errors.  }
         \end{table}

\begin{figure}
\centering
\resizebox{7cm}{!}{\includegraphics{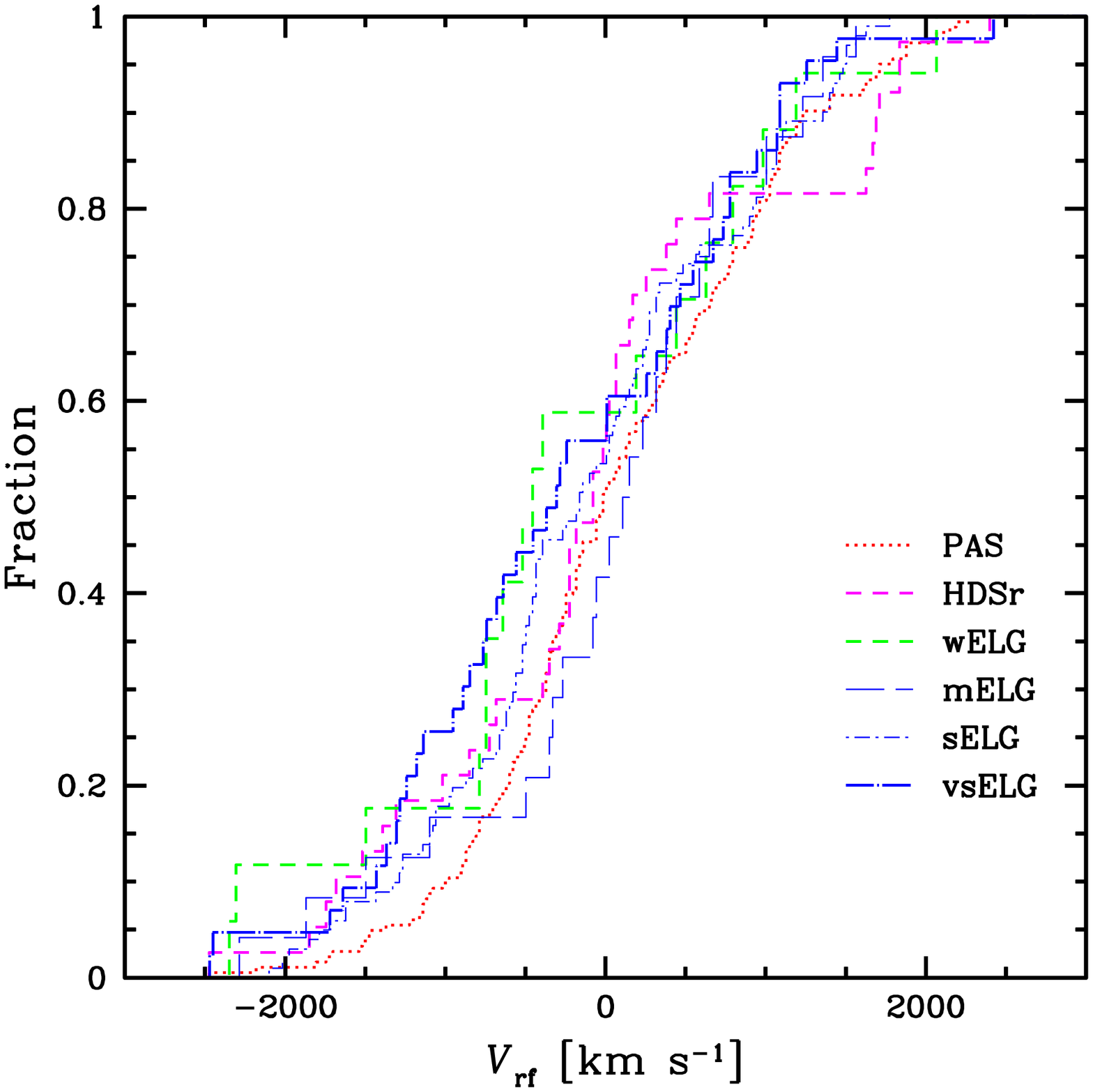}}
\resizebox{7cm}{!}{\includegraphics{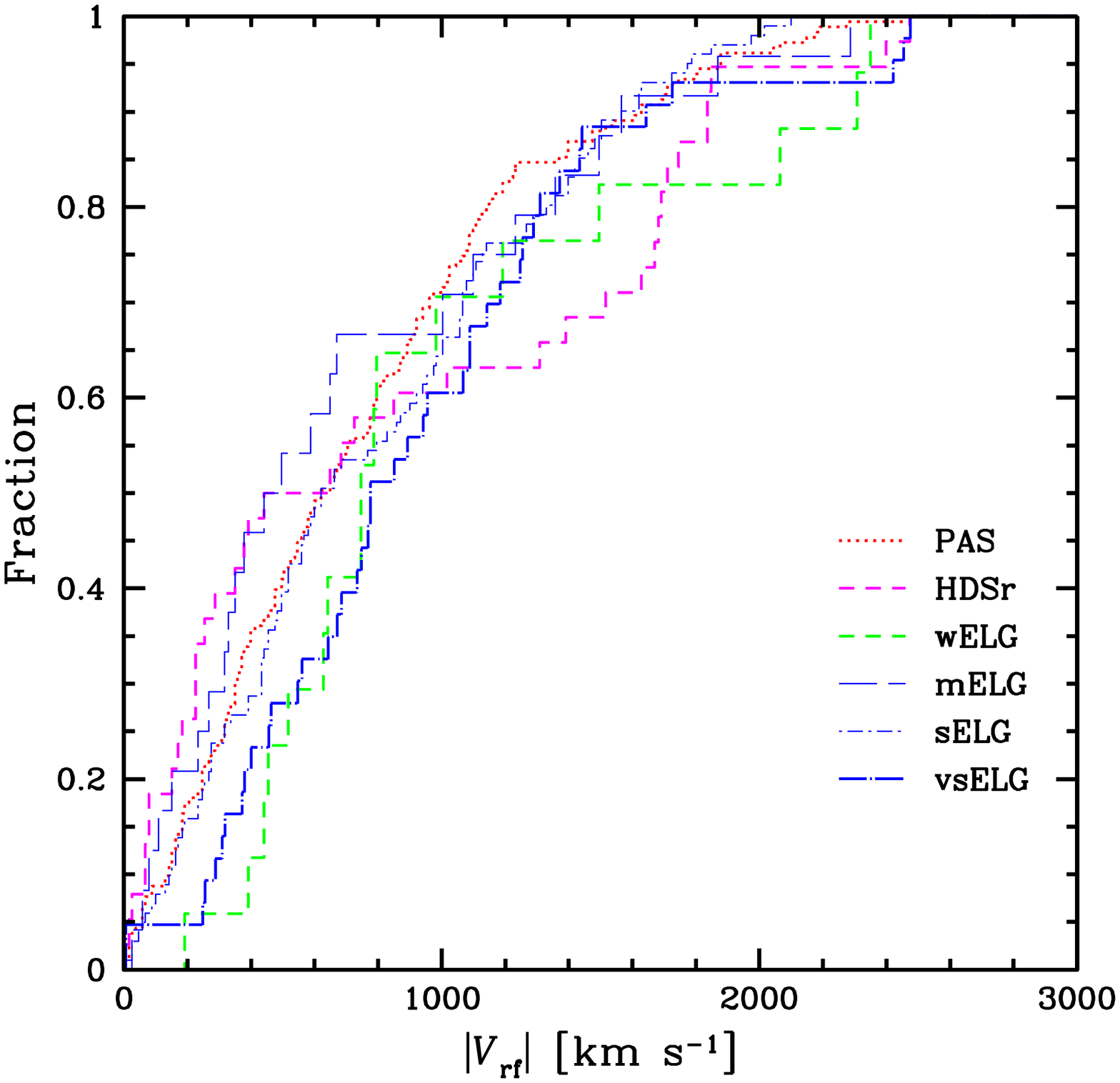}}
\caption
{Cumulative distributions of rest-frame LOS velocity $V_{\rm rf}$ 
and absolute $V_{\rm rf}$ of galaxies per spectral class
({\em upper} and {\em lower panels}, respectively).
}
\label{figmatkwov}
\end{figure}

The application of the KW-test to the rest-frame velocities $V_{\rm
  rf}$ [i.e., $(V-\left<V\right>)/(1+z)$] gives no significant result,
i.e., the velocity distributions of different spectral classes can
originate from the same parent distribution (see
Fig.~\ref{figmatkwov}, upper panel).  We also applied the KW-test to
the absolute value of $V_{\rm rf}$ in such a way to be more
specifically sensitive to differences in $\sigma_V$ values: no
significant result was obtained (see Fig.~\ref{figmatkwov}, lower
panel). When comparing two by two the $\left<V\right>$ estimates of
different spectral classes (see Table~\ref{tabv}), we found no
significant difference.  The same is true for $\sigma_V$.  Our results
corroborate previous studies claiming for a velocity distribution
which does not depend dramatically on galaxy color or spectral type
(Rines et al. \cite{rin05}; Mahajan et al. \cite{mah11}; Rines et
al. \cite{rin13}), including the results of B13 on MACS1206 itself. On
the other hand, several previous studies have found significant
differences in the velocity distributions of different galaxy
populations, the velocity dispersion of the population of blue SF
galaxies being higher than that of the population of red passive
galaxies (e.g., Tammann \cite{tam72}, Moss \& Dickens \cite{mos77},
Sodr\'e et al. \cite{sod89}, Biviano et al. \cite{biv92}, Zabludoff \&
Franx \cite{zab93}, Colless \& Dunn \cite{col96}; Biviano et
al. \cite{biv97}, Adami et al. \cite{ada98}, Dressler et al.
\cite{dre99}). The cause for this controversy may lie in i) the
selection criteria of member galaxies of past studies, which were
often not based on extensive spectroscopic information; ii) the
  dynamical status of the analyzed clusters; or iii) an evolutionary trend
  when considering clusters at different redshifts.  We defer a more
  complete analysis of this question when our analysis has been  extended
to other CLASH-VLT clusters. The specific case of MACS1206 is
  further analyzed at the end of this section.

\begin{figure}
\centering
\resizebox{12.7cm}{!}{\includegraphics{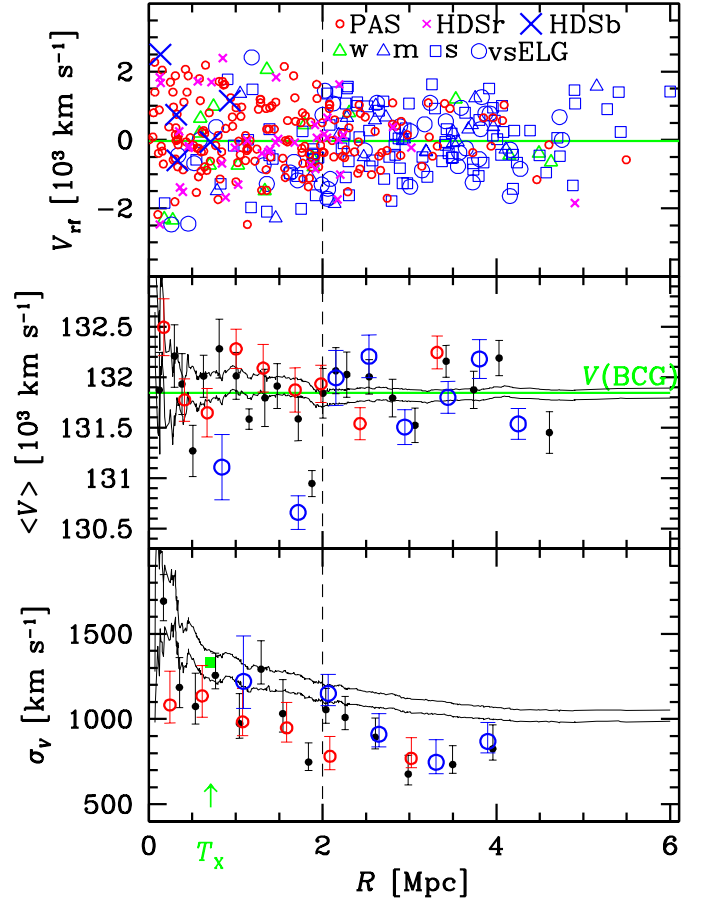}}
\caption
{{\em Top panel:} rest-frame LOS velocity vs projected clustercentric
  distance for galaxies of different spectral classes.  The cluster
  center coincides with the position of the BCG and the vertical
  dashed line indicates the value of $R_{200}$.  The horizontal green
  line indicates the BCG velocity.  {\em Middle panel.}  Integral
  profile of the mean LOS velocity shown through its error bands at
  the $68\%$ c.l..  The mean velocity at a given projected radius
  from the cluster center is estimated by considering all galaxies
  within that radius. Small black  dots, small red circles, and blue
  circles show the differential profiles for TOT, PAS, and ELG
  galaxies (each point is the value computed using 20 galaxies).  {\em
    Bottom panel.} As in the middle panel, but for the LOS velocity
  dispersion (here each point is the value computed using 30
  galaxies).  The green square indicates the estimate of the X-ray
  temperature $T_{\rm X}$ reported by Postman et al. (\cite{pos12})
  using {\em Chandra} data within the $R=0.714$ \h radius, indicated
  by the position of the green arrow, and here converted to $\sigma_V$
  assuming the density-energy equipartition between ICM and galaxies,
  i.e.,  $\beta_{\rm spec}=1$ (see Sect.~\ref{relax}).  }
\label{figvdspec}
\end{figure}

The spatial separation between PAS and ELG galaxies is also clear in
Fig.~\ref{figvdspec} (top panel), where we show the distribution of
galaxies of different spectral classes in the plot of rest-frame velocity
vs projected clustercentric distance, the so-called projected
phase-space diagram. In the same figure we also plot the integral and
differential $\left<V\right>$ and $\sigma_V$ profiles (middle and
bottom panels).  The integral $\left<V\right>$ profile shows that the
measure of the mean velocity is independent of the limiting radius of
the sampled region in agreement with a scenario of a unimodal, relaxed
cluster.  As for the differential $\left<V\right>$ profiles, the most
noticeable feature is the small value of $\left<V\right>$ of the ELG
and TOT populations at $R\siml 2$ \hh. This feature is likely related
to the presence of a low-velocity substructure detected at $\siml
R_{200}$ in the NE quadrant (see Sect.~\ref{3D}).

The  integral and differential $\sigma_V$ profiles show that the
velocity dispersion decreases with $R$.  This trend is the result of
the cluster mass density profile and the anisotropy velocity profile
as discussed in B13.  The interesting feature is that the face values
of $\sigma_V$ as computed for the ELG population are higher than those
of the PAS population at comparable radii. This is also shown in
Fig.~3 of B13, where the two populations are color defined. However,
the errors in individual bins are large and the difference is scarcely
significant.  To further investigate this point we used two
  approaches.  We fitted the velocity dispersion profiles of ELG and
  PAS galaxies as given in Fig.~\ref{figvdspec} (bottom panel)
  obtaining ${\rm lg}(\sigma_{V,{\rm
      ELG}})=3.15(\pm0.124)-0.399(\pm0.135)\times{\rm lg}(R)$ and
  ${\rm lg}(\sigma_{V,{\rm
      PAS}})=2.98(\pm0.121)-0.161(\pm0.047)\times{\rm lg}(R)$, where
  $\sigma_{V}$ and $R$ are in units of \ks and \hh.  The slopes are
  different at the 92\% c.l.  according to the Welch test (e.g., Guest
  \cite{gue61}). In the second approach we considered the region
  between 1 and 3 \hh, where both populations are well represented,
  obtaining $\sigma_{V,{\rm ELG,1-3}}=1085_{-68}^{+67}$ for 99 ELG
  galaxies and $\sigma_{V,{\rm PAS,1-3}}=857_{-58}^{+56}$ for 82 PAS
  galaxies, i.e., $\sigma_{V,{\rm ELG,1-3}}> \sigma_{V,{\rm
      PAS,1-3}}$ at the 97\% c.l. according to the F-test (see, e.g.,
  Press et al. \cite{pre92}).  We conclude that, in the case of
  MACS1206, the PAS and ELG populations do differ in their velocity
  dispersion but this difference is hidden by the combination of the
  trend of the $\sigma_V$ profile with the strong PAS vs ELG spatial
  segregation when global values are analyzed (see the first part of
  this section).

\section{Cluster substructure}
\label{sub}

We applied a set of tests in the velocity space (1D-tests), in the 2D
space of positions projected on the sky, and in the combined 2D+1D
space (3D-tests).  Since the sensitivity of individual diagnostics
depends on the relative position and velocity of substructure, no
single substructure test is the most sensitive in all situations
(Pinkney et al.  \cite{pin96}).  

\subsection{Analysis of the velocity distribution}
\label{1D}

The velocity distribution was analyzed to search for possible
deviations from Gaussianity that might provide important signatures of
complex internal dynamics.  We used the two robust shape estimators by
Bird \& Beers (\cite{bir93}),  the asymmetry index (AI) and the
tail index (TI). Following the Indicator test by Gebhardt \& Beers
(\cite{geb91}), we also checked the peculiarity
of the BCG velocity ($V_{\rm BCG}=131\,849$ \kss) in all the samples
containing the BCG (namely the TOT, R200, and PAS samples).  

\begin{figure}
\centering
\resizebox{\hsize}{!}{\includegraphics{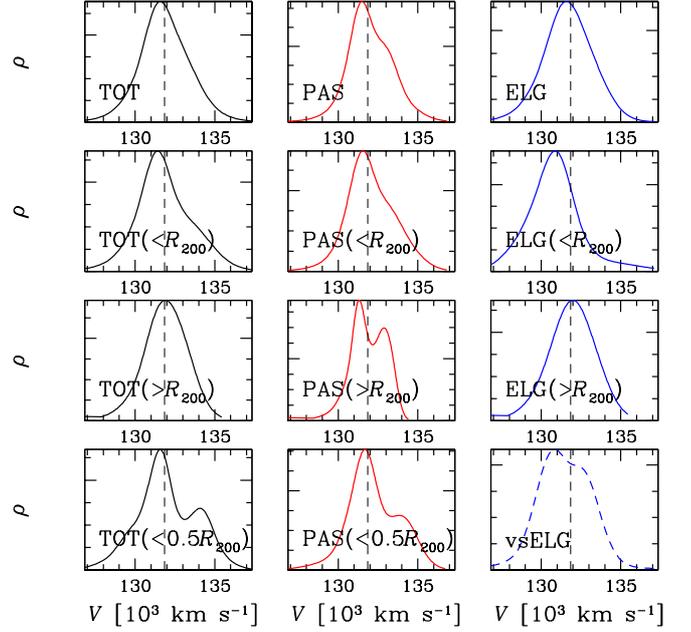}}
\caption
{Galaxy density reconstruction in the LOS velocity space through the
  1D-DEDICA method.  Results for all galaxies in the TOT, PAS, and ELG
  samples are shown from left to right, while results for the whole
  and partial cluster regions are shown from top to bottom.  The result
  in the vsELG sample is shown in the bottom-right panel. In each
  panel, the dashed vertical line indicates the BCG velocity.  Units on
  the y-axis are rescaled to the maximum value of each density distribution.  }
\label{figk1mulv}
\end{figure}

To detect and analyze possible departures from a single-peak
distribution, we used the 1D-DEDICA method, already applied in
  Sect.~\ref{cat} to determine MACS1206 membership.  DEDICA was 
  introduced by Pisani (\cite{pis93}), and extended to the
  multivariate case in Pisani (\cite{pis96}). The algorithm is
  described in the original studies and references therein. Here we
  summarize useful points (see also  Appendix A in Girardi et al.
  \cite{gir96}).  DEDICA is a method of cluster analysis based on the
  estimate of the probability density of a data sample, where the
  density is estimated by using an iterative and adaptive kernel
  estimator, in this case a Gaussian kernel with an optimal choice of
  the width. It is based on the assumption that a cluster corresponds
  to a local maximum in the density of points. This method allows us
  to estimate i) the statistical significance of each subcluster it
  detects and ii) the probability that each galaxy in the data sample is
  a member of each detected subcluster. In particular, it gives a measure
  of the overlapping between two systems (see Girardi et
  al. \cite{gir96} and Pisani et al. \cite{pis96}). We note that
  DEDICA is a non-parametric method in the sense that it does not
  require any assumption about the number of clusters or any other 
  of their features.  Table~\ref{tabsub} summarizes the results of
  the 1D-DEDICA method applied to the TOT and other samples of
  MACS1206 and Fig.~\ref{figk1mulv} shows the relevant plots.

\begin{table*}
        \caption[]{Results of the substructure analysis.}
         \label{tabsub}
                $$
         \begin{array}{|l r |  c  c c c | c | c l l|}
            \hline
            \noalign{\smallskip}
            \hline
            \noalign{\smallskip}
&
&\multicolumn{4}{c|}{\mathrm{1D}} 
&\multicolumn{1}{c|}{\mathrm{2D}}
&\multicolumn{3}{c|}{\mathrm{2D}+\mathrm{1D}}\\
\mathrm{Sample} & N_{\rm g} &
\mathrm{TI}&\mathrm{AI}& \mathrm{V_{\mathrm{BCG}}pec.}& \mathrm{DED.}&\mathrm{DED.}& 
\mathrm{Vgrad.}&\mathrm{DS}\left<V\right>&\mathrm{DS}\sigma_{V,{\rm corr}}
\ (\mathrm{DS}\sigma_{V})\\
& 
&\%&\%&\%&\mathrm{N_p}&
\mathrm{N_p}&
\%& \% & \phantom{AAA}\% \\
            \hline
            \noalign{\smallskip}
\mathrm{WHOLE}  &445&ns   &ns   &ns&1           &1+3          &ns&ns  &\phantom{AAA}ns\ (>99.9)\\
\mathrm{R200 }  &249&ns   &ns   &ns&1           &1+1          &ns&ns  &\phantom{AAA}ns\ (ns)\\
\mathrm{PAS  }  &183&90-95&ns   &ns&1           &1+1          &ns&ns  &\phantom{AAA}94\ (92)\\
\mathrm{HDSr }  & 38&90-95&95-99& -&1           &2            &ns&ns  &\phantom{AAA}ns\ (ns)\\
\mathrm{wELG}   &17 &ns   &ns   & -&-           &1            &ns&-&\phantom{AAA}-\\
\mathrm{ELG}    &168&90-95&ns   & -&1           &7            &ns&>99.9&\phantom{AAA}ns\ (99.7)\\
\mathrm{vsELG}  & 43&ns   &ns   & -&1           &2            &ns&99.1&\phantom{AAA}ns\ (94)\\
              \noalign{\smallskip}
            \hline
         \end{array}
$$ \tablefoot{Column~1: sample ID; Col.~2: number of galaxies, $N_{\rm
             g}$; Cols.~3 and 4: significance of the deviations from
           Gaussian according to the tail and asymmetry indices; Col.~5:
           significance of the peculiarity of the BCG velocity;
           Col.~6: number of peaks detected through the
           1D-DEDICA method, only peaks detected with a c.l. larger
           than $99\%$ are considered;  Col.~7: number of peaks
           detected through the 2D-DEDICA method, where n1+n2 
           indicates the presence of n1 major peaks and n2 minor with
            very low-density peaks; Col.~8: significance of the
           existence of a velocity gradient; Cols.~9 and
           10: significance of the existence of substructure according
           to the DS$\left<V\right>$-test and the 
           DS$\sigma_V$-test, the latter for both the
           profile-corrected and standard versions. Only significance values larger than $90\%$ are reported, while non-significant values are indicated with ns.}
         \end{table*}

The velocity distributions of the TOT and R200
samples show no departures from a Gaussian, nor signs of a peculiar
BCG velocity, and only one significant peak is detected by the
1D-DEDICA reconstruction.  Within 0.5$R_{200}$, two peaks, formed of
91 and 37 galaxies, are detected by the 1D-DEDICA reconstruction (see
Fig.~\ref{figk1mulv}). However, the two groups are  strongly
overlapped with 26$+$26 galaxies having a non-negligible probability
to belong to both groups and not different in their 2D distributions.

The velocity distributions of galaxy populations per spectral class
show  marginal or no departures from a Gaussian with the exception of
the HDSr population. In the case of the PAS class no sign of a
peculiar BCG velocity is found. Moreover, only one significant peak is
detected by the 1D-DEDICA reconstruction with the exception of wELG, a
very poor class where a non-significant peak can be detected.  The
velocity distribution of the PAS galaxies outside $R_{200}$ and that
of the PAS galaxies within 0.5$R_{200}$ show the presence of two
peaks, but, as in the case discussed above, the two groups are
strongly overlapped and show no difference in their spatial
distribution.

We note the peculiar velocity distribution of ELGs 
within $R_{200}$, which is significantly peaked at lower values with
respect to the mean velocity of the whole cluster, $\left<V_{\rm
  ELG,<R200}\right>=(130\,709\pm$171) \kss. The velocity distribution
of vsELGs is peaked around the same value and, in addition, the visual
inspection of the 1D-DEDICA reconstruction (Fig.~\ref{figk1mulv},
bottom-right panel) suggests a secondary bump at high
velocities.  These features are related to the presence of a low-velocity NE galaxy structure and of a high-velocity SEext galaxy
structure as we discuss in Sect.~\ref{3D}.  We do not plot the
reconstructions for HDSr and wELG, whose distributions are poorly sampled.

\subsection{Analysis of the 2D galaxy distribution}
\label{2D}

\begin{figure}
\centering
\resizebox{\hsize}{!}{\includegraphics{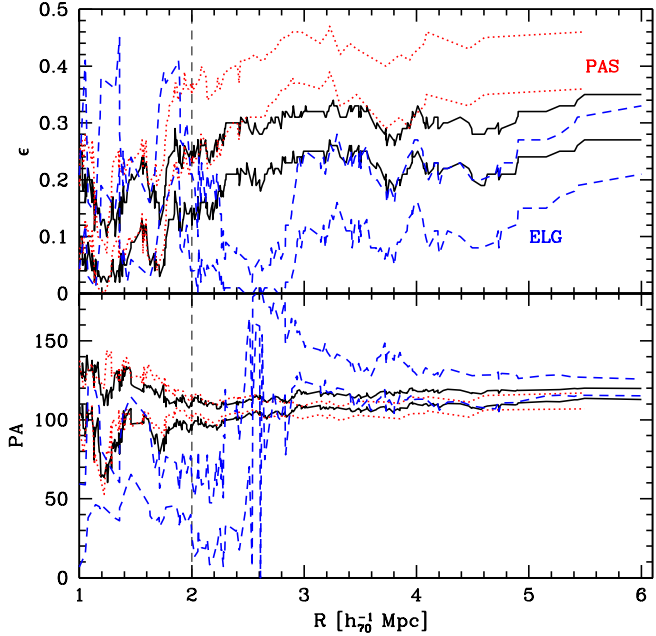}}
\caption
{Integral profiles of ellipticity ({\em upper panel}) and position
  angle ({\em lower panel}) for the whole galaxy population
  (solid black line) and per spectral class according to the labels in
  the upper panel.  The values of $\epsilon$ and PA at a given
  radius $R$ are estimated by considering all galaxies within
  $R$. Results within 1 \h are too noisy to be meaningful.
  The vertical dashed line indicates the value of $R_{200}$.}
\label{figell}
\end{figure}

The main features in Fig.~\ref{fig2dspec} are the WNW-ESE elongation
and the lack of ELGs in the SW quadrant with respect to the NE
quadrant (14 vs 54 galaxies).  In the computation of ellipticity
($\epsilon$) and position angle of the major axis (PA), we followed
the moments of inertia method (Carter \& Metcalfe \cite{car80}; see
also Plionis \& Basilakos \cite{pli02} with weight $w=1$).
Table~\ref{tabell} lists the values of $\epsilon$ and PA (measured
counterclockwise from north) for the TOT and R200 samples, and the
main spectral classes.  Figure~\ref{figell} shows the integral
estimates of $\epsilon$ and PA at increasing radii.  As far as
concerning PAS galaxies, the value of $\epsilon$ increases out to
$R\sim R_{200}$, which is the radius containing most PASs, and then is
stable. The value of PA is already stable at $R\mincir R_{200}$. As
for ELGs, Figure~\ref{figell} allows us to appreciate a strong
variation of $\epsilon$ and PA at $R\sim 2$-3$R_{200}$, while values
are roughly stable at $R\simg 3$ \hh. This variation is likely due to
the NE vs SW quadrant asymmetry and, specifically, to the presence of
an important NE structure as discussed at the end of the section.
When comparing values at large radii, $R\simg 3$ \hh, the PAS and ELG
spatial distributions have similar position angles,
PA$=110$-$120$\degreee, while the ELG distribution is somewhat rounder
than the PAS distribution, $\epsilon_{\rm PAS}\sim
1.5$-$2\epsilon_{\rm ELG}$ at the 2-$3\sigma$ c.l. depending on the
precise radius.  Table~\ref{tabell} also lists the values of
$\epsilon$ and PA for the HDSr population, but the respective profiles
are not shown in Fig.~\ref{figell} for the sake of clarity.  The value
of PA$_{\rm HDSr}$ agrees with that of PA$_{\rm PAS}$ within 1$\sigma$
c.l.. The value of $\epsilon_{\rm HDSr}$ is larger than $\epsilon_{\rm
  PAS}$ only at a $\siml 2\sigma$ c.l., the large value of
$\epsilon_{\rm HDSr}$ being due to the few galaxies of the dense ESE
peak (see in the following).

\begin{table}
        \caption[]{Ellipticity and position angle of the galaxy distribution.}
         \label{tabell}
                $$
         \begin{array}{l c c}
            \hline
            \noalign{\smallskip}
            \hline
            \noalign{\smallskip}
\mathrm{Sample} & \epsilon&\mathrm{PA}\\
                &         &\mathrm{deg}\\
            \hline
            \noalign{\smallskip}
\mathrm{TOT}  &0.31_{-0.04}^{+0.04}&117_{-4}^{+3}\\
\mathrm{R200 }  &0.20_{-0.06}^{+0.05}&102_{-8}^{+7}\\
\mathrm{PAS}      &0.41_{-0.05}^{+0.05}&111_{-4}^{+6}\\
\mathrm{HDSr }       &0.55_{-0.06}^{+0.07}&105_{-4}^{+8}\\
\mathrm{ELG }       &0.26_{-0.05}^{+0.07}&121_{-6}^{+5}\\
              \noalign{\smallskip}
            \hline
         \end{array}
$$
\end{table}

\begin{figure}
\centering
\resizebox{\hsize}{!}{\includegraphics{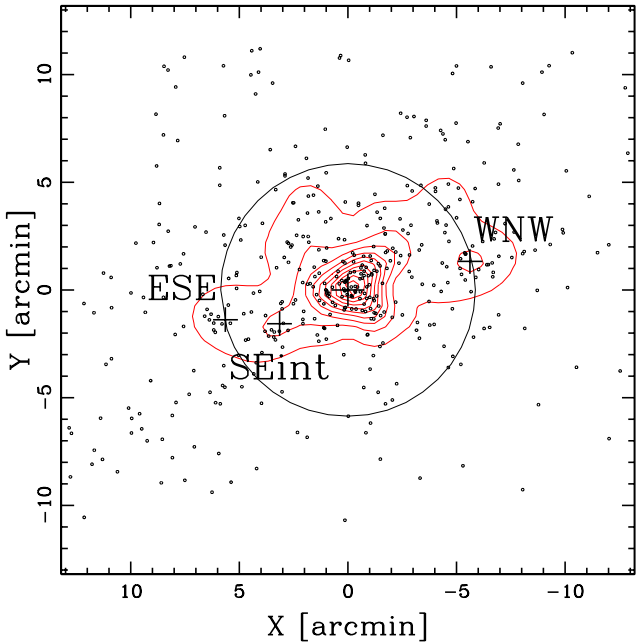}}
\caption{Spatial distribution and isodensity contours of the 445
  cluster members (TOT sample). Large cross and circle highlight the
  BCG position and the $R_{200}$ region. Small crosses indicate the
  secondary density peaks (see Table~\ref{tabdedica2d}).}
\label{figk2z}
\end{figure}

\begin{figure*}
\centering
\includegraphics[width=9cm]{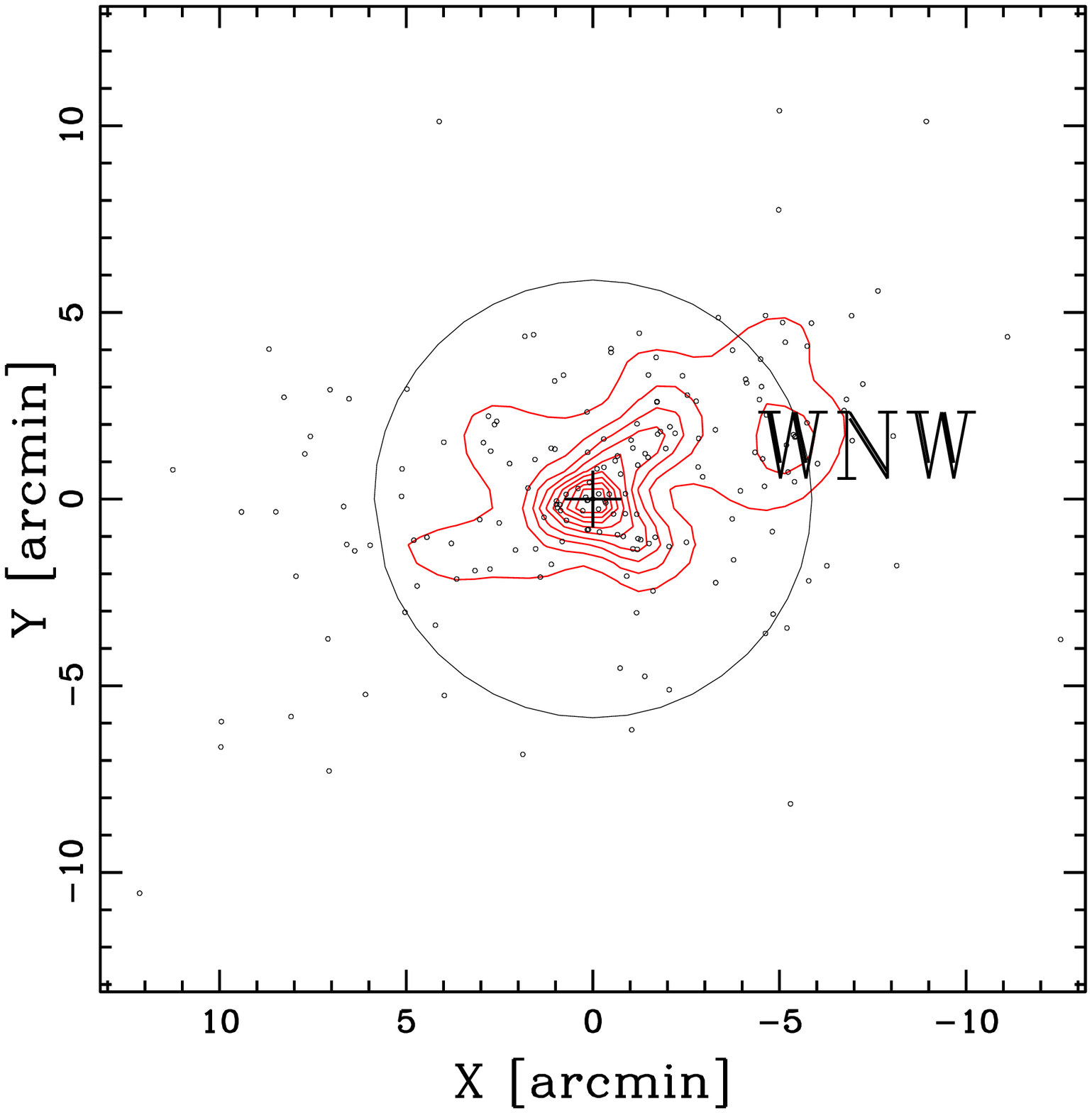}
\includegraphics[width=9cm]{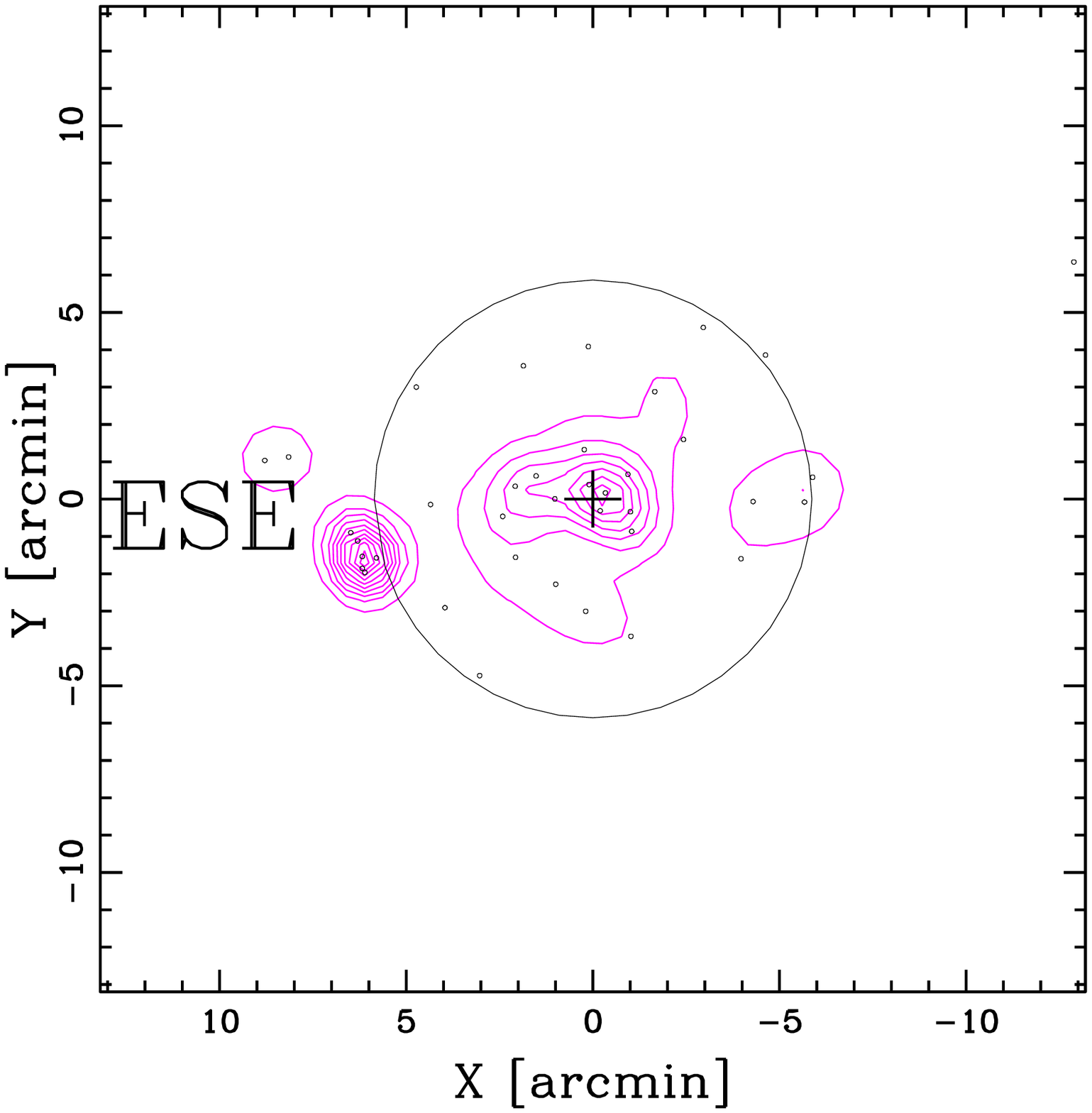}
\includegraphics[width=9cm]{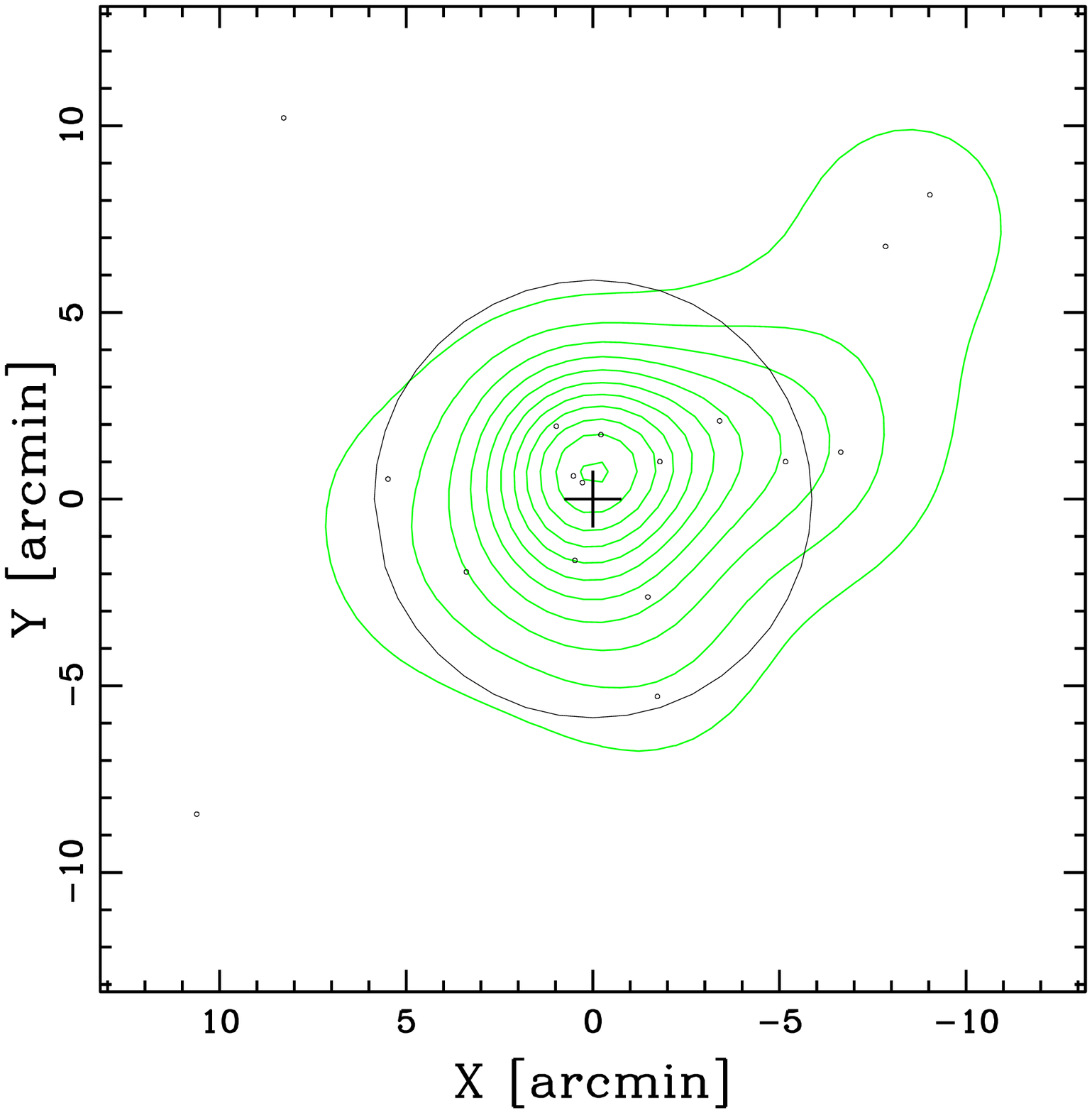}
\includegraphics[width=9cm]{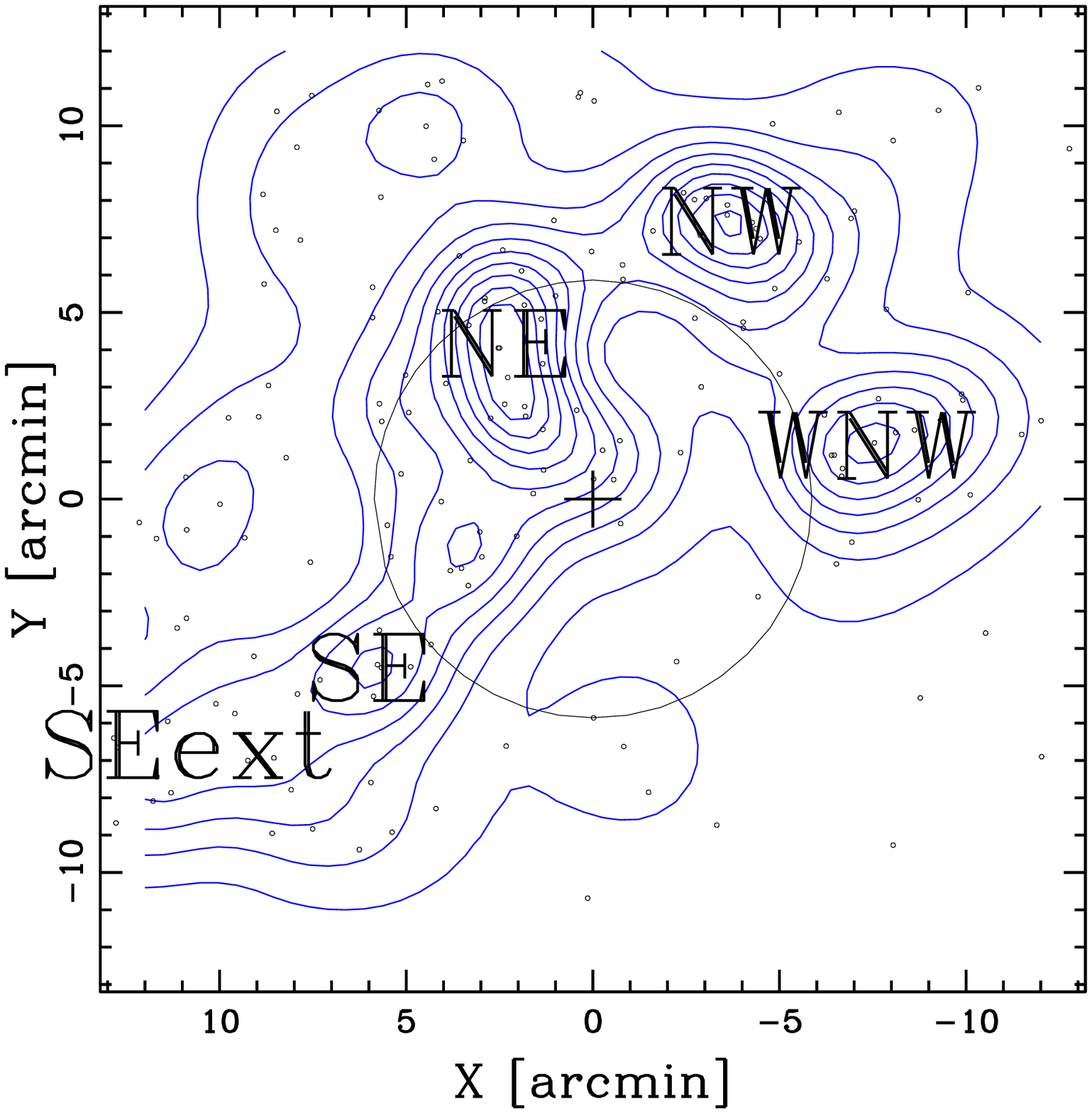}
\caption{Spatial distributions and isodensity contour maps for the PAS
  ({\em upper left  panel}), HDSr (upper {\em right  panel}), wELG (lower {\em
    left  panel}), and ELG (lower {\em right  panel}) populations.
  Labels refer to the peaks listed in Table~\ref{tabdedica2d}.  The
  $R_{200}$ region is highlighted by the circle centered on the BCG.}
\label{figk2zmul}
\end{figure*}

We also analyzed the galaxy spatial distribution through the 2D-DEDICA
analysis (Pisani \cite{pis96}).  In Table~\ref{tabdedica2d}, we
present the full information for the relevant peaks, i.e., those
with a c.l.$\geq 99\%$, with a relative density with respect to the
main peak $\rho_{\rm S}\simg 0.20$, and with at least 10 assigned
galaxies.  The relevant maps are shown in Figs.~\ref{figk2z} and
\ref{figk2zmul}.  When analyzing  the whole cluster, the 2D-DEDICA density
reconstruction confirms the cluster elongation and shows a main
structure peaked on the BCG with a few secondary peaks (at ESE, SE,
and WNW). The SE peak in the internal cluster region (hereafter
SEint), with 38 assigned galaxies, is the only secondary peak detected
in the R200 sample.

\begin{table}
        \caption[]{Results of the 2D-DEDICA analysis.}
         \label{tabdedica2d}
            $$
         \begin{array}{l r c c r }
            \hline
            \noalign{\smallskip}
            \hline
            \noalign{\smallskip}
\mathrm{Subclump} & N_{\rm S} & \alpha({\rm J}2000),\,\delta({\rm J}2000)&\rho_{\rm S}&\chi^2_{\rm S}\\
& & \mathrm{h:m:s,\degree:\arcmm:\arcs}&&\\
         \hline
         \noalign{\smallskip}
\mathrm{TOT-main}      &219&12\ 06\ 11.5-08\ 47\ 57&1.00&103\\
\mathrm{TOT-WNW}       &102&12\ 05\ 49.4-08\ 46\ 44&0.20& 29\\
\mathrm{TOT-SEint}     & 38&12\ 06\ 24.9-08\ 49\ 37&0.20& 16\\
\mathrm{(TOT-ESE)}     & 41&12\ 06\ 35.0-08\ 49\ 26&0.15& 15\\
         \hline
\mathrm{R200-main}       &203&12\ 06\ 11.5-08\ 47\ 57&1.00& 80\\
\mathrm{R200-SEint}      & 30&12\ 06\ 24.8-08\ 49\ 40&0.20& 17\\
         \hline
\mathrm{PAS-main}        &126&12\ 06\ 12.2-08\ 48\ 05&1.00& 41\\
\mathrm{PAS-WNW}         & 44&12\ 05\ 51.0-08\ 46\ 30&0.25& 12\\
         \hline
\mathrm{(HDSr-ESE)}      &  6&12\ 06\ 37.0-08\ 49\ 37&1.00& 6\\
\mathrm{HDSr-main}       & 15&12\ 06\ 11.0-08\ 47\ 57&0.67& 4\\
         \hline
\mathrm{wELG}            & 15&12\ 06\ 12.0-08\ 47\ 20&1.00& 7\\
         \hline
\mathrm{ELG-NE}          & 39&12\ 06\ 21.8-08\ 44\ 00&1.00&14\\
\mathrm{ELG-NW}          & 30&12\ 05\ 57.2-08\ 40\ 38&0.85&11\\
\mathrm{ELG-WNW}         & 24&12\ 05\ 42.0-08\ 46\ 34&0.68& 9\\
\mathrm{ELG-SE}          & 13&12\ 06\ 36.0-08\ 52\ 34&0.58& 5\\
\mathrm{ELG-SEext}       & 20&12\ 06\ 49.7-08\ 54\ 19&0.43&10\\
\mathrm{ELG-NEext}       & 17&12\ 06\ 31.0-08\ 38\ 10&0.31& 8\\
\mathrm{ELG-ESEext}      & 10&12\ 06\ 54.2-08\ 48\ 28&0.31& 5\\
         \hline
\mathrm{vsELG-NE}       & 24&12\ 06\ 20.5-08\ 44\ 20&1.00&9\\
\mathrm{vsELG-SEext}    & 15&12\ 06\ 49.3-08\ 53\ 13&0.55&4\\
              \noalign{\smallskip}
              \noalign{\smallskip}
            \hline
         \end{array}
$$ \tablefoot{ Column.~1: subsample/peak ID, labels in parentheses
           correspond to galaxy peaks that do not completely fulfil
           the criteria we fixed for galaxy number or density, but
           are listed here becasue they are  discussed in the text;  Col.~2: number of
           assigned member galaxies, $N_{\rm S}$; Col.~3: R.A. and Dec
           of the density peak; Col.~4: relative density with respect
           to the highest peak, $\rho_{\rm S}$; Col.~5: $\chi^2$ value
           of the peak. For each population, the number of peaks is
           summarized in Table~\ref{tabsub} (Col.~7).  }
\end{table}

\begin{figure}
\centering
\resizebox{7cm}{!}{\includegraphics{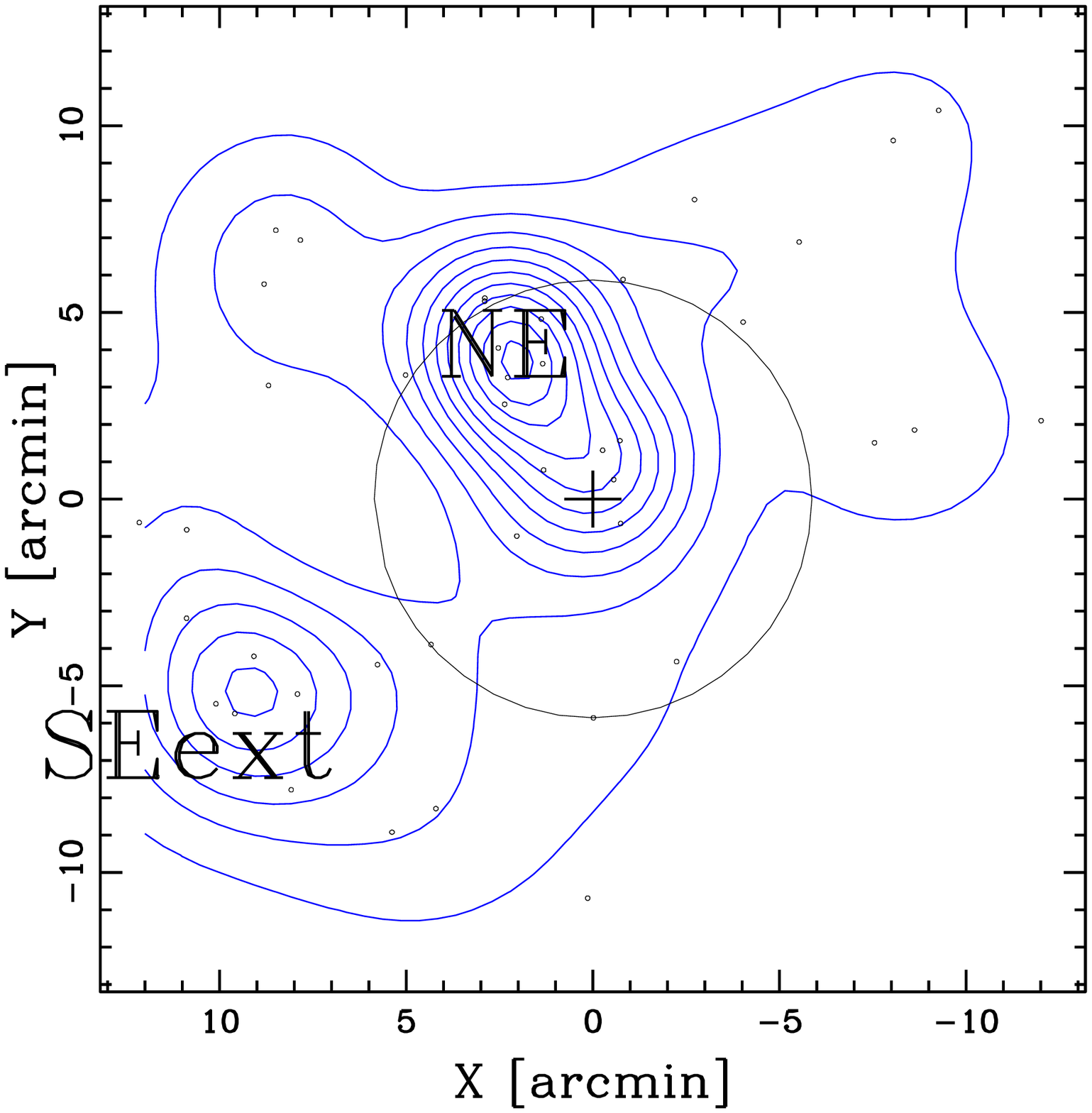}}
\centering
\resizebox{7cm}{!}{\includegraphics{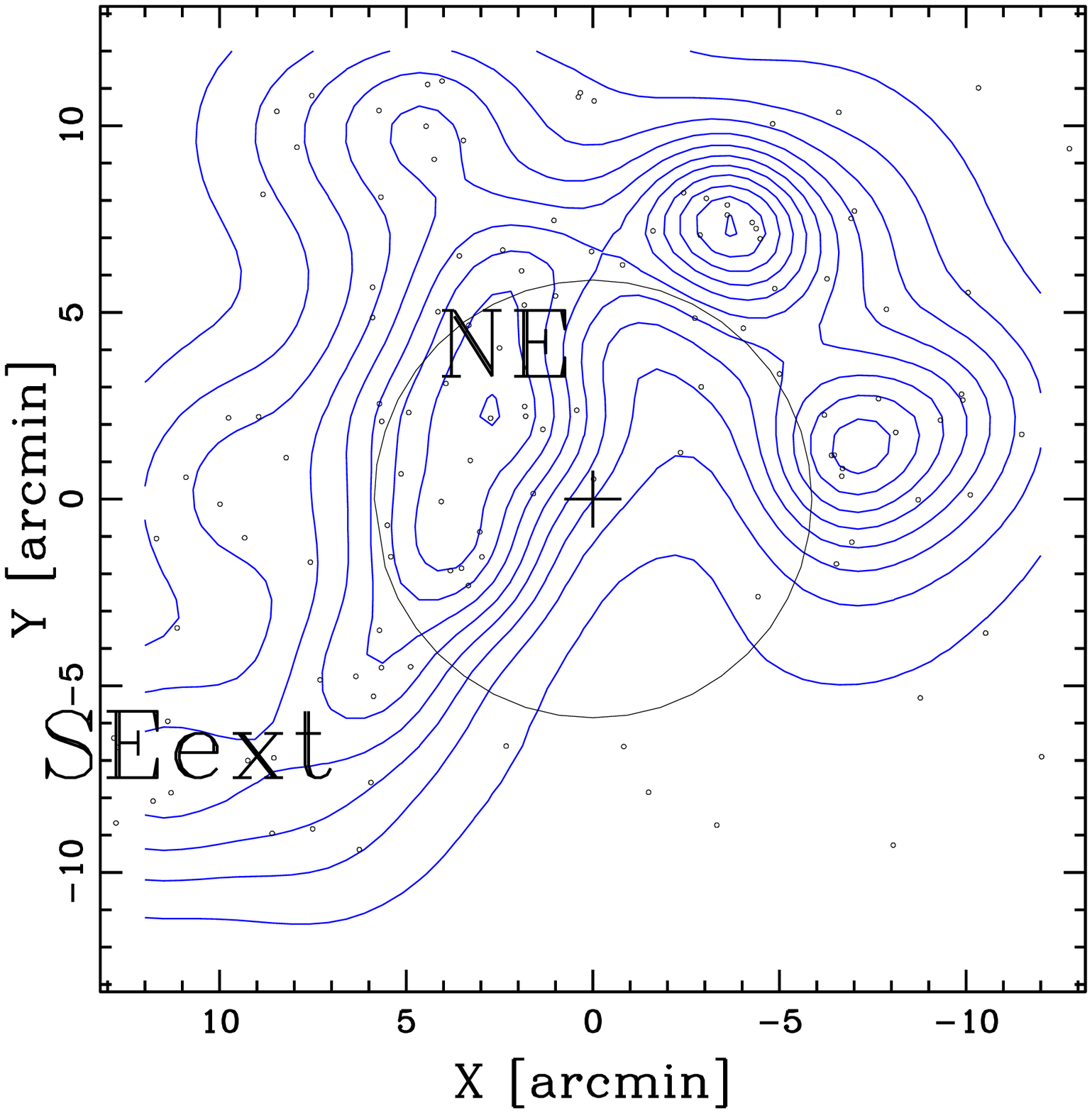}}
\caption{Spatial distribution and isodensity contour maps of vsELGs
  ({\em upper panel}) and other ELGs (i.e., mELGs$+$sELGs, {\em lower
  panel}). The position of labels are those in Fig.~\ref{figk2zmul}
(lower right  panel) to allow an easy comparison with that figure and
highlight the concentration of vsELGs in the NE and SEext structures.
The $R_{200}$ region is indicated by the circle centered on the BCG.
}
\label{figk2zELGs40}
\end{figure}

When  considering PAS galaxies, the main feature is the clear
elongation along the WNW-ESE direction in the plane of the sky, with
the presence of an external secondary peak at WNW.  Noticeably, HDSr
galaxies trace a very dense peak $\sim 2$ \h at ESE, denser than that
around the cluster center.  This ESE peak is formed of only six
galaxies, but its existence is supported by our independent analysis
in Sect.~\ref{col} based on galaxies of different colors.  We
detected no peculiarity in the velocity field in the ESE region.  The
2D-DEDICA analysis of the wELG population shows only one peak roughly
centered on the cluster center.  To the contrary, the 2D distribution
of ELGs is clearly not radially symmetric, with a lack of galaxies in
the SW region and shows several clumps (see
Table~\ref{tabdedica2d}). The ELG distribution shown in
Fig.~\ref{figk2zmul} allows us to better interpret the variation of
$\epsilon$ and PA in Fig.~\ref{figell}.  The NE structure centered at
$R\siml R_{200}$ is the cause of the value of PA$\sim 50$\degree
at $R \sim R_{200}$.  When more external ELG structures, somewhat related
to the NE-SW direction, are included the PA value shows a fast,
strong change and the $\epsilon$ value also increases.
According to our analysis in Sect.~\ref{pop}, the spatial distribution
of vsELGs is different from that of mELGs and sELGs.  The reason is
that vsELGs are concentrated in the NE and SEext structures, as shown
by the contour maps for vsELGs and mELGs$+$sELG galaxies (see
Fig.~\ref{figk2zELGs40}).

The study of Pisani (\cite{pis96}) shows that the 2D-DEDICA
  method, based on the centrally symmetric two dimensional Gaussian
  kernel, can reconstruct the density of an elongated 2D system even
  in the case with an axial ratio $a/b=0.1$, although slightly rounder
  isoplethes are obtained (see his Fig.~2). Pisani
  (\cite{pis96}) also shows that a simulated cluster with $a/b=0.1$
and  strongly contaminated by a population of randomly distributed points
  (100 random points and 100 cluster points) is split into a series of clusters
  strongly overlapping each other along the major axis (see his Fig.4
  and Table 1).  This second case is very extreme with respect to
  MACS1206 both for the presence of an important background and the
  high ellipticity (cf. $a/b=0.1$ and  $a/b=0.7$).  However, to leave
  no room for doubt in the case of MACS1206, we run an additional
  test.  We performed the circularization of the coordinates in the
  TOT sample, leaving  the cluster center fixed and using the
  ellipticity listed in Table~\ref{tabell}, and repeated the full
  2D-DEDICA analysis.  We confirmed the presence and the relative density
  of the main  WNW and ESE peaks, but not the SEint peak.  However,
  the SEint peak and its relative density are confirmed in the
  re-analysis of the R200 sample after the circularization procedure.
  The WNW and ESE peaks are also independently confirmed in the
  re-analysis of the PAS and HDSr samples.

\subsection{Combining position and velocity information}
\label{3D}

The existence of correlations between positions and velocities of
cluster galaxies is always a strong footprint of real substructures.
To study the 3D cluster structure we applied two tests.  The presence
of a velocity gradient was searched for by  performing a multiple linear
regression fit to the observed velocities with respect to the galaxy
positions in the plane of the sky (e.g., den Hartog \& Katgert
\cite{den96}, Girardi et al. \cite{gir96}).  The significance is based
on 1000 Monte Carlo simulated clusters obtained by shuffling galaxy
velocities with respect to their positions. We found no significant
evidence for a velocity gradient.

The $\Delta$-statistics devised by Dressler \& Schectman
(\cite{dre88}, hereafter DS-test) is a powerful test for 3D
substructure, which is valid in samples down to 30 member galaxies,
at least in the case of major mergers (Pinkney et al. \cite{pin96}).
For each galaxy, the deviation $\delta_i$ is defined as $\delta_i^2 =
[(N_{\rm{nn}}+1)/\sigma_{\rm{V}}^2][(\left<V\right>_{\rm loc}
  -\left<V\right>)^2+(\sigma_{V,{\rm loc}} - \sigma_V)^2]$, where the
subscript loc denotes the local quantities computed over the
$N_{\rm{nn}}=10$ neighbors of the galaxy,  $\left<V\right>$ and
$\sigma_V$ are the global quantities, and $\Delta$ is the sum of the
$|\delta_i|$ of the individual $N$ galaxies of the sample.  The
significance of $\Delta$, i.e., of the existence of substructure, is
checked by running 1000 Monte Carlo simulations, randomly shuffling
the galaxy velocities. Here, we used two kinematical estimators
alternative to the $\delta_{i}$ parameter.  We considered separately
the contribution of the deviation of the local mean velocity from
global mean velocity $\delta_{i,V}= [(N_{\rm
    nn}+1)^{1/2}/\sigma_V]\times (\left<V\right>_{\rm loc}
-\left<V\right>)$ and the contribution of the deviation of the local
velocity dispersion from the global velocity dispersion $\delta_{i,
  {\rm s}}= [(N_{\rm nn}+1)^{1/2}/\sigma_V]\times (\sigma_{V,{\rm
    loc}} - \sigma_V)$ (Barrena al. \cite{bar11}; see also Girardi et
al. \cite{gir97}).  The results are listed in Table~\ref{tabsub}.

\begin{figure}
\centering 
\resizebox{\hsize}{!}{\includegraphics{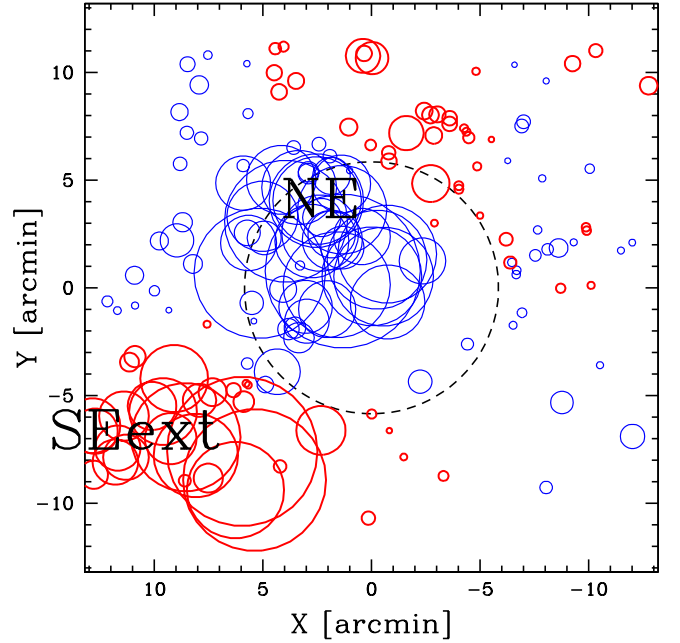}}
\caption
 {DS$\left<V\right>$ bubble-plot for the ELG population. The larger
   the circle, the larger  is the deviation of the local mean velocity
   from the global mean velocity.  Blue and heavy red circles show
   where the local value of mean velocity is smaller or larger than
   the global value.  Labels indicate the two peaks detected in the 2D
   analysis (see Table~\ref{tabdedica2d} and Fig.~\ref{figk2zmul},
   lower right  panel).  The $R_{200}$ region is highlighted by the
   large black dashed circle centered on the BCG.}
\label{figdssegno10vELG}
\end{figure}

\begin{figure}
\centering 
\resizebox{\hsize}{!}{\includegraphics{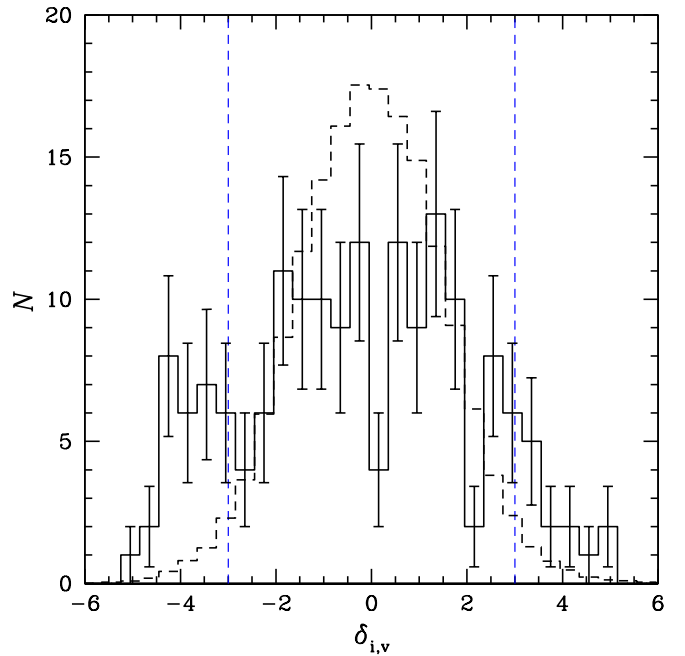}}
\caption
{The distribution  of the $\delta_{i,V}$ values
for the ELG population. The solid line histogram indicates observed
galaxies. The dashed histogram indicates the galaxies of
  simulated samples, normalized to the observed galaxy number. The blue
  vertical dashed lines highlight the $|\delta_{i,V}|>3$ regions where most
  galaxies are expected to belong to the 3D substructure.}
\label{figdeltai}
\end{figure}

The DS$\left<V\right>$-test returns a strong positive detection of
substructure in the case of the ELG population ($>99.9\%$ c.l.), with
a low-velocity group within $R_{200}$ at NE and a high-velocity group
in the external SE region (see Fig.~\ref{figdssegno10vELG} for the
so-called bubble plot).  To be more quantitative, we resorted to the
technique developed by Biviano et al. (\cite{biv02}, see also Girardi
et al. \cite{gir06}). We compared the distribution of the $\delta_{i,
  V}$ values of the real ELG galaxies to the distribution of the
$\delta_{i, V}$ values of the galaxies of all the 1000 Monte Carlo
simulated ELG samples (Fig.~\ref{figdeltai}).  In agreement with the
result of the DS$\left<V\right>$-test, the two distributions differ at
the $>99.99\%$ c.l. according to the 1DKS-test.  The distribution of
the values of real galaxies shows a tail at large positive and
negative $\delta_{i, V}$ values, populated by the galaxies belonging
to the 3D substructure. In particular, the low-velocity tail is more
populated than the high-velocity tail in agreement with the relative
importance of the low-velocity NE substructure in
Fig.~\ref{figdssegno10vELG}.  To quantify the number of galaxies
involved in the 3D substructure, we assumed a threshold of
$|\delta_{i, V}|= 3$ and counted 46 real galaxies with $|\delta_{i,
  V}|> 3$, while five galaxies are predicted by simulations; in other words,
$\sim 41$ ELGs are expected to belong to the 3D substructure.  The
galaxies with $|\delta_{i, V}|> 3$ are spatially separated in the NE
and SE clumps allowing us to estimate the rest-frame velocity of these
clumps, $\Delta V_{\rm rf,NE}\sim -550$ \kss and $\Delta V_{\rm
  rf,SEext}\sim +550$ \kss.

\begin{figure}
\centering 
\resizebox{\hsize}{!}{\includegraphics{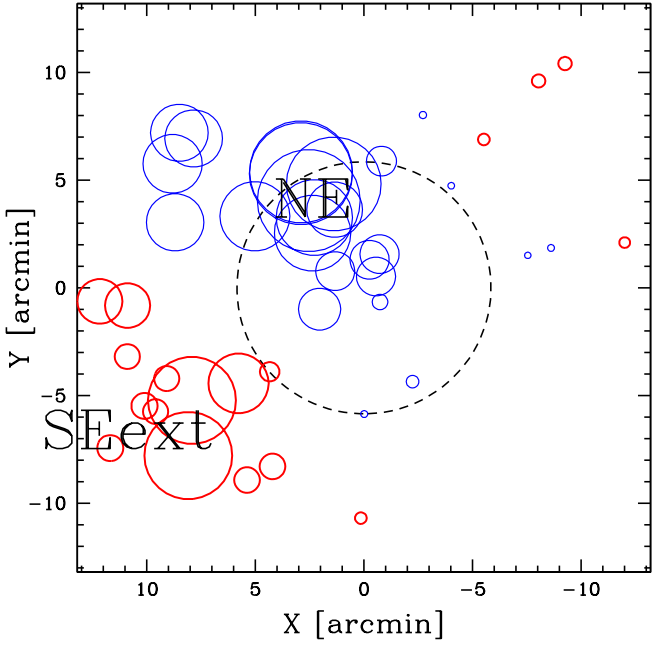}}
\caption
 {DS$\left<V\right>$ bubble-plot as in Fig.~\ref{figdssegno10vELG} but
   for vsELGs. 
}
\label{figdssegno10vELGs40}
\end{figure}

We note that the tails of the ELG $\delta_{i,V}$ distribution are
populated by several  vsELGs; 20 vsELGs, i.e., about half of the whole
vsELG population, have $|\delta_{i, V}|> 2.5$.  This prompted us to
apply the DS$\left<V\right>$-test to the ELG subclasses.  The test applied
to the vsELG class returns, in spite of the small number of vsELG galaxies, a
$99.1\%$ c.l. signal of substructure (see
Fig.~\ref{figdssegno10vELGs40}), while no signal is found for mELG and
sELG.  From these results and those of the previous section we can conclude
that vsELGs are strongly concentrated in the NE and SEext
substructures and are related to the 3D substructure.

In the context of the above results, the features in the velocity
distributions of ELGs and vsELGs plotted in Fig.~\ref{figk1mulv} can
be explained. The velocity distribution of ELGs within $R_{200}$
is peaked at low velocity owing to the low-velocity NE substructure,
very important within $R_{200}$ (see Fig.~\ref{figdssegno10vELG}).  In
the case of vsELGs (whole sample), the high-velocity SEext
substructure is also important resulting in a velocity distribution
suggestive of  two peaks.

\begin{figure}
\centering 
\resizebox{7cm}{!}{\includegraphics{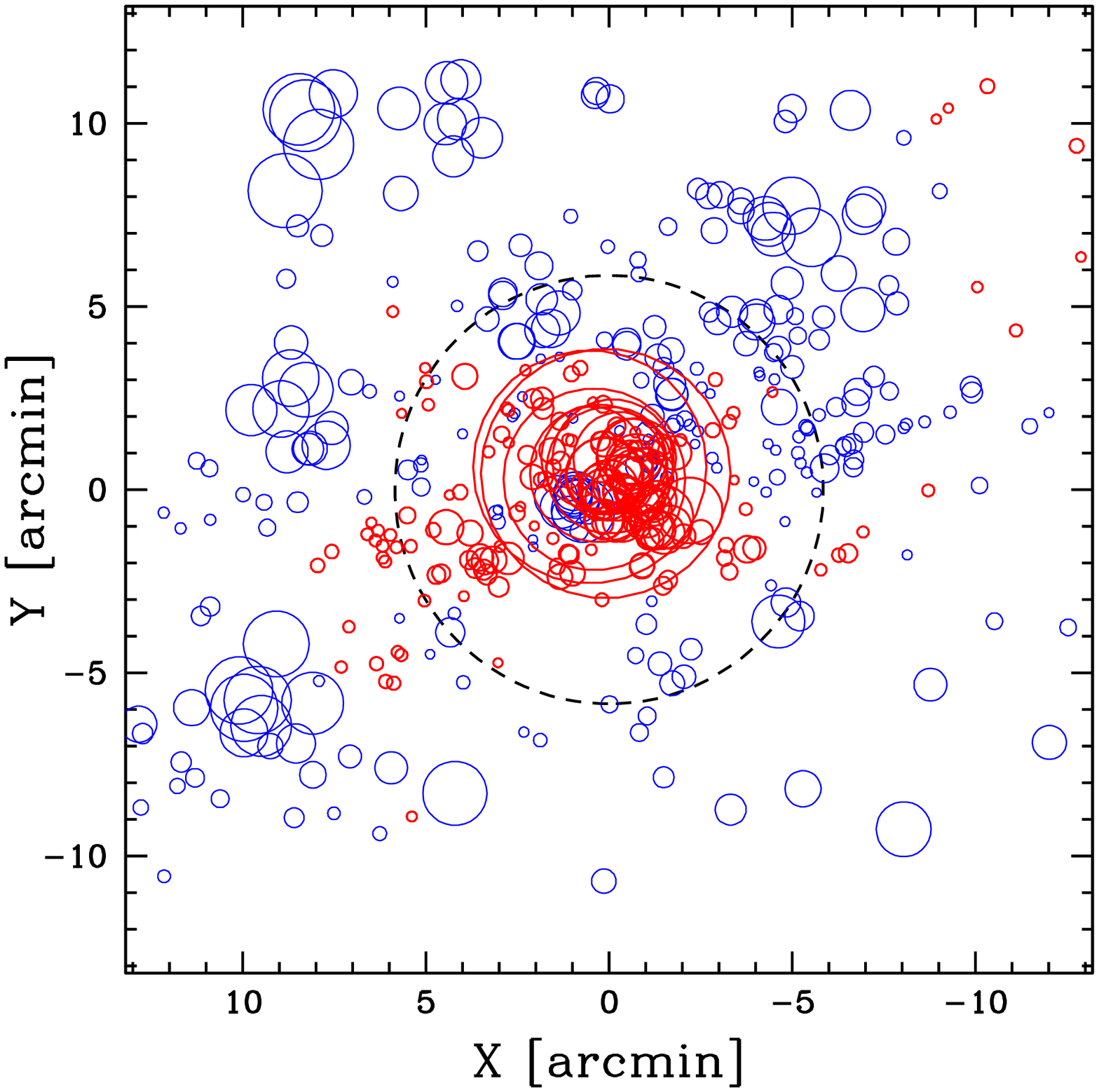}}
\resizebox{7cm}{!}{\includegraphics{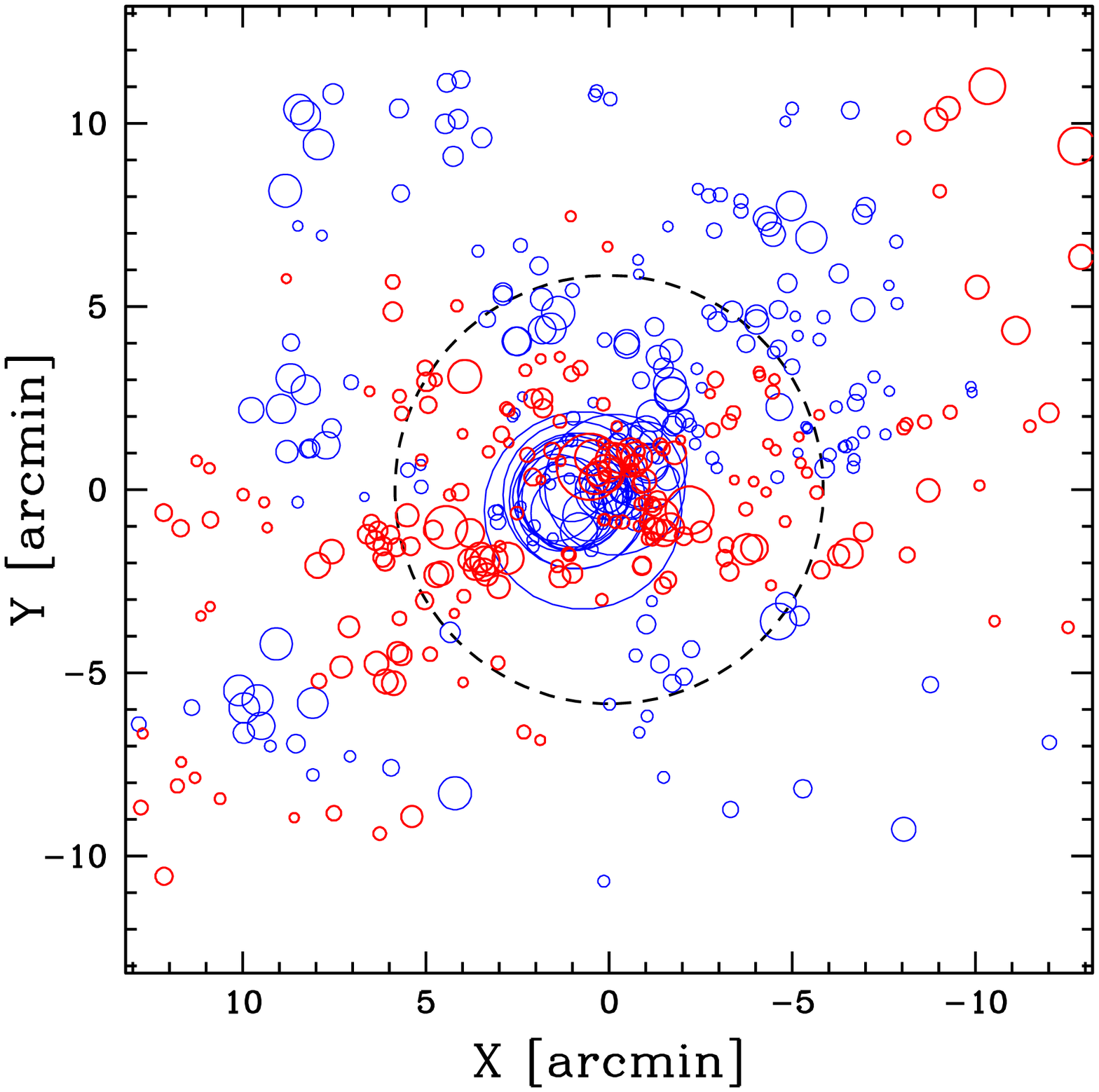}}
\caption
 {DS$\sigma_V$ bubble-plot in the standard and profile corrected cases
   for the TOT sample ({\em upper} and {\em lower panels}).  The
   larger the circle, the larger is the deviation of the local
   velocity dispersion from the global velocity dispersion.  Blue and
   heavy red circles show where the local value is smaller or larger
   than the global value.  In the {\em upper panel} the signal of high
   local $\sigma_V$ in the central region and low local $\sigma_V$ in
   the external region reflect the trend of the $\sigma_V$ profile
   (see Fig.~\ref{figvdspec}, bottom panel) and is no longer
   significant when applying the suitable correction ({\em lower
     panel}).  The $R_{200}$ region is indicated by the dashed
   circle centered on the BCG.  }
\label{figdssegno10s}
\end{figure}

As for the DS$\sigma_V$-test, the MACS1206 dataset is so large in
galaxy content and in sampled area that the radial $\sigma_V$ profile
(Fig.~\ref{figvdspec}, bottom panel; see also Fig.~3 in B13) should
be taken into account.  The application of the standard
DS$\sigma_V$-test in the TOT sample detects a spurious strong signal
of substructure ($>99.9\%$ in the TOT sample) with a peak of high
local velocity dispersion in the central cluster region and low values
of local velocity dispersion in external cluster regions (see
Fig.~\ref{figdssegno10s}).  We overcame this difficulty by correcting the
kinematical estimator.  Specifically, we consider the deviation of
$\sigma_{V,{\rm loc}}$ from the velocity dispersion expected at the
corresponding radius $R$, i.e., $\delta_{i, {\rm s,corr}}= [(N_{\rm
    nn}+1)^{1/2}/\sigma_V] \times (\sigma_{V,{\rm loc}} -
\sigma_{V,{\rm R}})$. The value of $\sigma_{V,{\rm R}}$ was obtained by
fitting the values of the $\sigma_{V}$-profile in Fig.~\ref{figvdspec}
[lg($\sigma_{V,{\rm R}}$/\kss)$=3.027-0.224\times$lg($R$/\hh)]. This
correction cancels the spurious signal of substructure in the TOT and
other samples (cf.  $\mathrm{DS}\sigma_{V,{\rm corr}}$ and
$\mathrm{DS}\sigma_V$ values in Table~\ref{tabsub} and cf. upper and
lower panels in Fig.~\ref{figdssegno10s}). We note that no significant
signal is obtained in the R200
sample, which is by definition less spatially extended,  even before the correction.

\begin{figure}
\centering 
\resizebox{7cm}{!}{\includegraphics{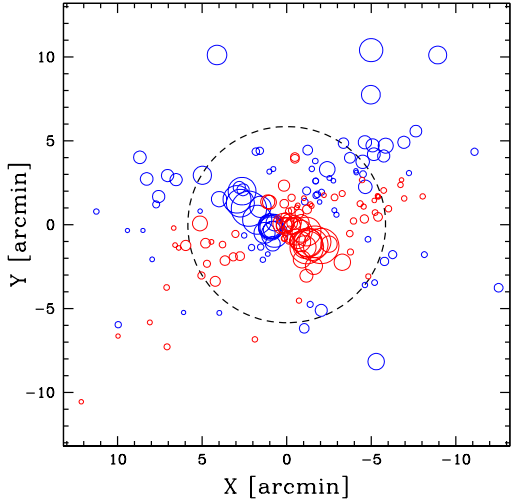}}
\caption
 {DS$\sigma_V$ bubble-plot as in Fig.~\ref{figdssegno10s} (standard
   case), but for the PAS population.}
\label{figdssegno10sP}
\end{figure}

As for the PAS sample, the DS$\sigma_V$-test substructure signal is
not radially symmetric (see Fig.~\ref{figdssegno10sP}) and, as a
consequence, does not vanish when applying the correction based on the
$\sigma_{V}$ profile.  However, the evidence of substructure is always
marginal (at the 92-$94\%$ c.l., see Table~\ref{tabsub}) and few galaxies are
related to this 3D substructure.  We do not discuss  this small
feature further.

All the above results are obtained using the classical version of DS-test
local groups with $N_{\rm nn}=10$ neighbors. We also applied the
version with $N_{\rm nn}=20$ for the samples having many members (TOT,
R200, PAS, and ELG) with no relevant difference in the results.

\subsection{Substructure analysis as a function of galaxy colors}
\label{col}

\begin{figure}
\centering 
\includegraphics[width=4.3cm]{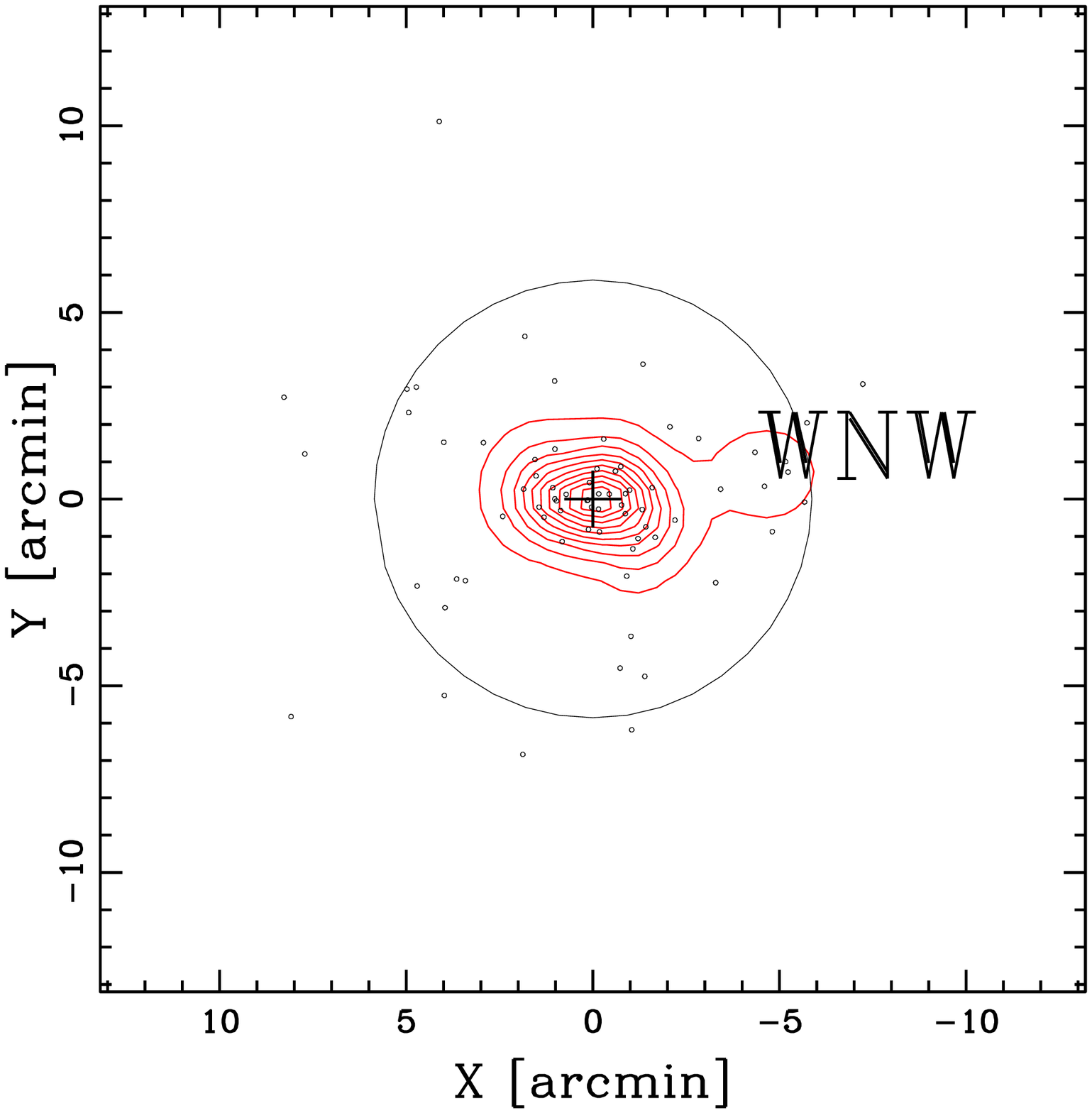}
\includegraphics[width=4.3cm]{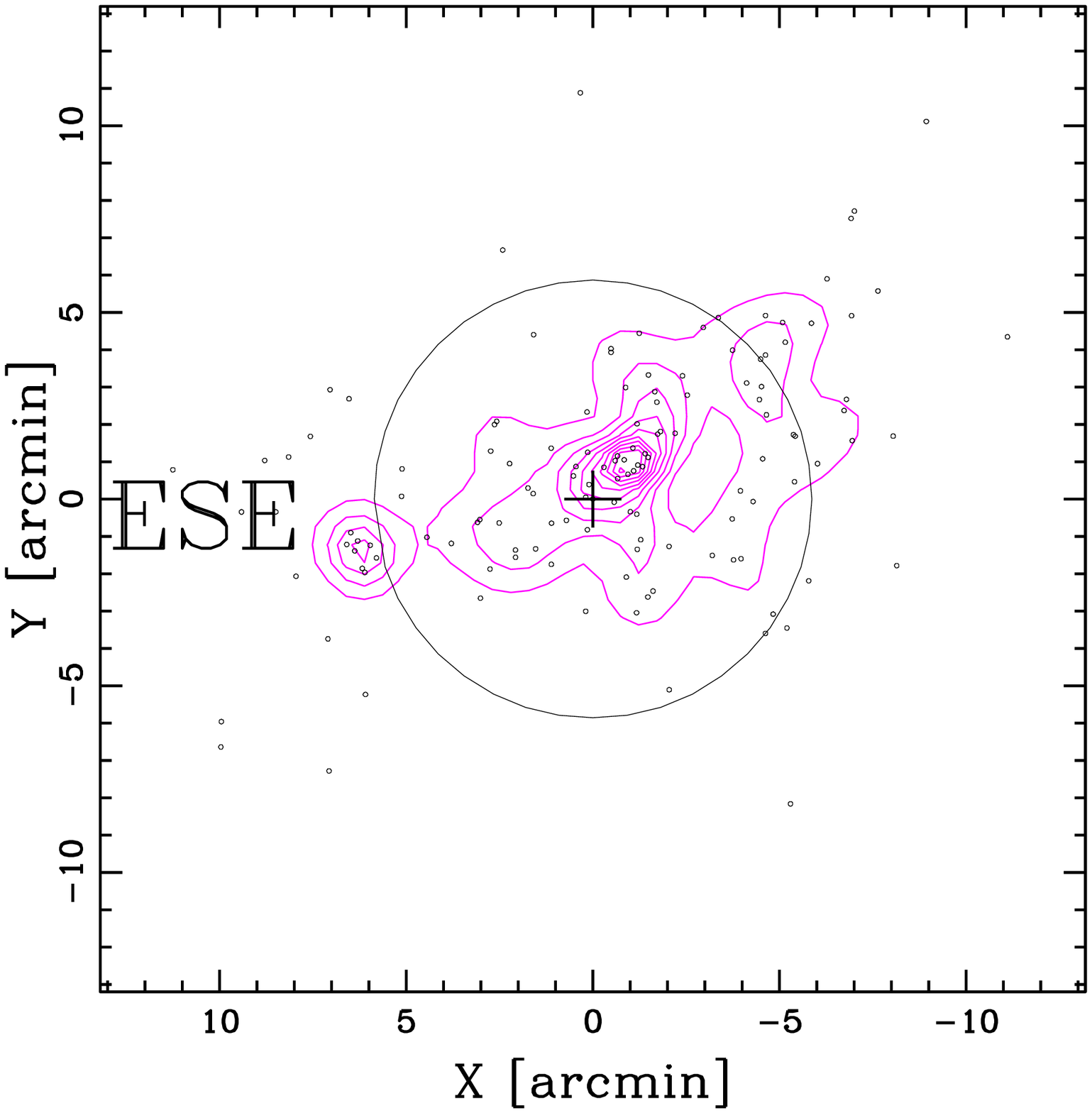}
\includegraphics[width=4.3cm]{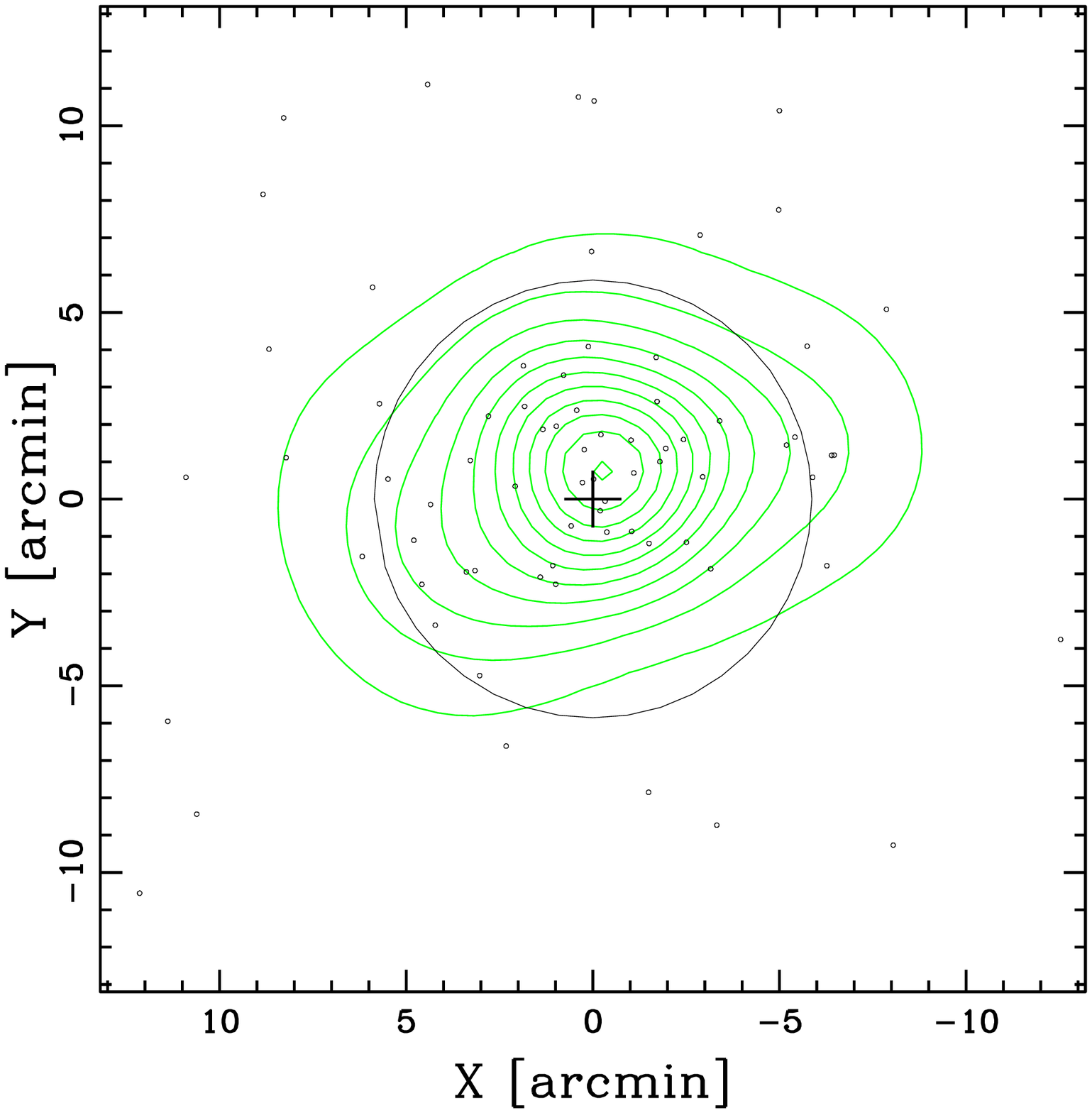}
\includegraphics[width=4.3cm]{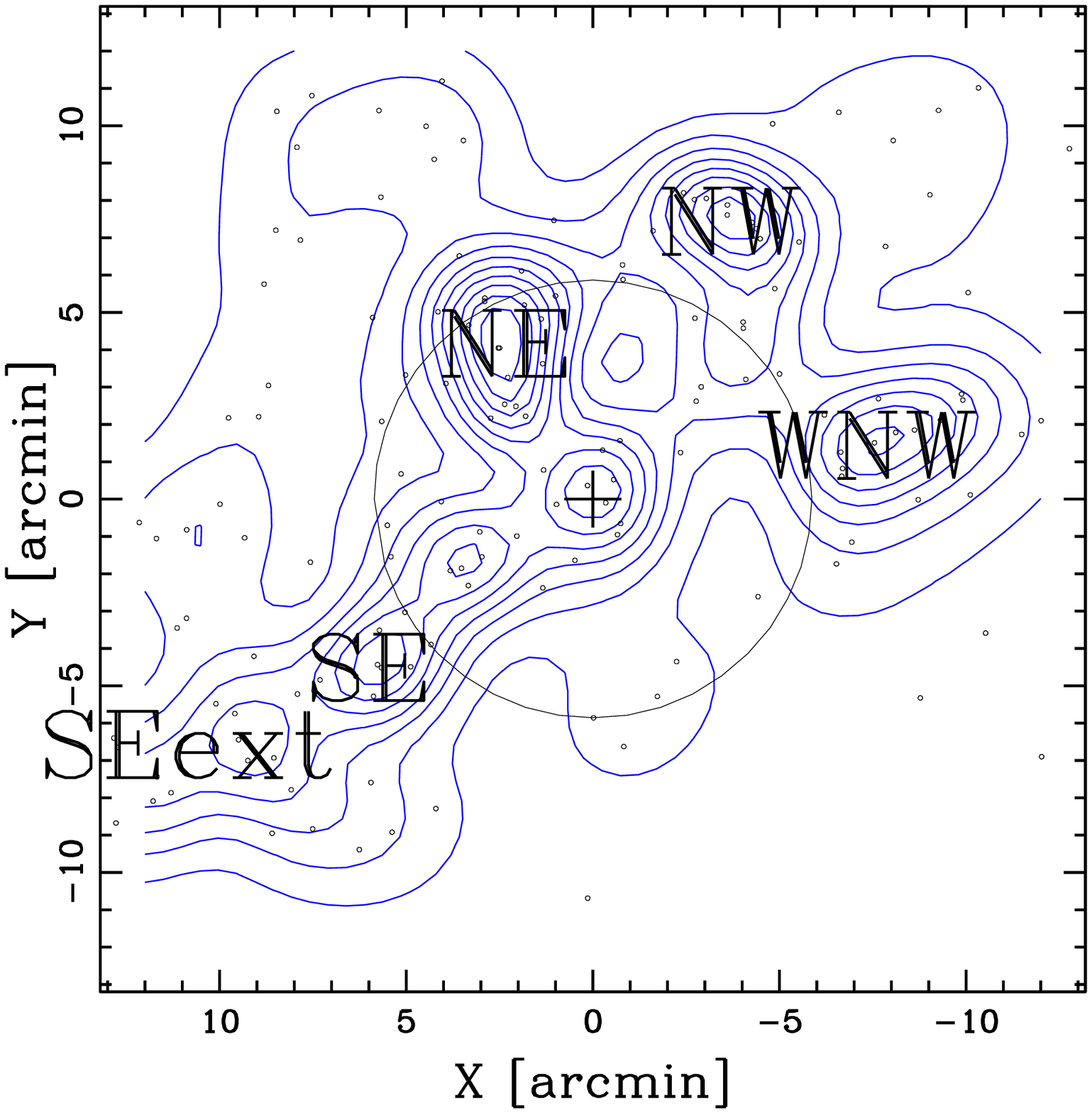}
\caption {Spatial distributions and isodensity contour maps for the
  four color classes: RedU ({\em  upper left panel}), RedD ({\em 
    upper right panel}), Green ({\em lower left panel}), and Blue ({\em
    lower right panel}) populations.  Labels are the same as in
  Fig.~\ref{figk2zmul}.}
\label{figk2zcol}
\end{figure}

We present here the main results of our substructure analysis of
MACS1206 by binning the sample on the basis of galaxy colors. We
divided the sample into four classes.  Starting from the red sequence
galaxies, i.e., those with $|(B-R_{\rm C})_{\rm diff}|\le 0.31$
\footnote{We adopt a threshold of  0.31 instead of 0.3 because we chose to include the BCG in the red sequence.}, we defined
the red sequence galaxies with positive and negative $(B-R_{\rm
  C})_{\rm diff}$, called the RedU  sample (72 galaxies) and the RedD sample (133
galaxies), respectively.  We then defined a Blue sample, containing 152
galaxies with $(B-R_{\rm C})_{\rm diff}<-0.5$, and at intermediate
colors the Green sample, including 74 galaxies with $-0.5<(B-R_{\rm
  C})_{\rm diff}<-0.31$. By repeating the aforementioned analysis per
color class, we found no strong evidence of non-Gaussianity in the
velocity distribution (i.e., always $<95\%$ c.l.), and no evidence of
multimodal distributions according to the 1D-DEDICA method. We show
the results of the 2D analysis in Fig.~\ref{figk2zcol}. The 2D density
maps for Green and Blue galaxies resemble those of wELGs and ELGs in
Fig.~\ref{figk2zmul}.  RedU galaxies are strongly clustered around the
cluster center; the WNW peak has a density $\rho_{\rm S}$ well under
the threshold we fixed as being relevant and it is displayed for
completeness.  RedD galaxies trace the elongated structure of the
cluster and, interestingly, confirm the presence of the ESE peak
(with 14 objects).

\section{General results and discussion}
\label{disc}

\subsection{Cluster structure}
\label{relax}

The  statistical results of the substructure tests applied to the TOT and
R200 samples indicate that MACS1206 does not show the presence of large-scale
substructure.

The 2D elongation is the most important large-scale feature in the
cluster structure.  In the plane of the sky the cluster structure appears
clearly elongated along the WNW-ESE direction (see
Fig.~\ref{figk2z}). We obtained  PA $= 102_{-8}^{+7}$ and
$117_{-4}^{+3}$ degrees for galaxies in the R200 and TOT samples,
respectively.  This orientation matches that of the mass
distribution from strong gravitational lensing (see Fig.~1 in Zitrin
et al.~\cite{zit12}), BCG shape, X-ray isophotes, and the mass
distribution from weak gravitational lensing (PA in the range
105-119\degreee, Table~6 in Umetsu et al.~\cite{ume12}), the 
asymmetric distribution of the intracluster light (PA of BCG and ICL
in the range 101-109\degreee, Presotto et al.~\cite{pre14}).  Our
novel result is that there is neither evidence for a velocity gradient
along the WNW-ESE direction nor for a velocity difference between the
main system and most subclumps. This suggests that the main MACS1206
structure is lying in the plane of the sky.

As for the interpretation, it is well known that cluster's dynamical
activity is strongly correlated with the tendency of clusters to be
elongated and aligned with the nearby clusters (Plionis \& Basilakos
\cite{pli02}) because of the anisotropic merging along the LSS
filaments as also shown by the analysis of simulations (e.g.,
Basilakos et al. \cite{bas06}; Cohen et al. \cite{coh14}). However, it
is also known that the final distribution after violent relaxation may
be aspherical (Aaarseth \& Binney \cite{aar78}; White \cite{whi96}).
Thus, in principle, the elongation of MACS1206 is not in contrast with
the lack of statistical evidence for the presence of important
substructure.

An interesting global parameter is the value of the ratio between
the energy per unit mass of galaxies to that of ICM as parametrized with
$\beta_{\rm spec}=\sigma_V^2/(kT_{\rm X}/\mu m_{\rm p})$, where
$T_{\rm X}$ is the X-ray temperature, $\mu=0.58$ is the mean molecular
weight and $m_{\rm p}$ the proton mass. The value $\beta_{\rm spec}=1$
indicates the density-energy equipartition between ICM and galaxies.
The mean $\beta_{\rm spec}$ value observed for massive clusters is
consistent with unity both in nearby (Girardi et al. \cite{gir96},
\cite{gir98}) and distant systems out to $z\sim 0.4$ (Mushotzky \&
Scharf \cite{mus97}), while large deviations may be related to
important cluster merging phenomena (e.g., Ishizaka~\cite{ish97}).  In the 
case of MACS1206, the value of $kT_{\rm X}=10.8\pm0.6$ keV has been obtained
by Postman et al.~(\cite{pos12}) using {\em Chandra} data within a
radius of 0.714 \h and excluding the very central cluster region.  Using our
estimate of $\sigma_V$ within the same radius, we obtained $\beta_{\rm
  spec}= 0.96\pm 0.14$, in agreement with the scenario of density-energy
equipartition (see also Fig.~\ref{figvdspec}, bottom panel).

At small scale, the analysis of the 2D galaxy distribution shows three
secondary peaks with low values of relative density $\rho_{\rm S}\siml 0.2$
with respect to the main cluster peak (WNW, SEint, ESE in the TOT
sample, see Table~\ref{tabdedica2d}). The ESE and WNW peaks have already been
 detected as significant overdensities by Lemze et
al. (\cite{lem13}), who have used a different sample and method.  In
the R200 sample, the SEint density peak is the only one detected,
leading to a $12\%$ of cluster members ($8\%$ in luminosity) related
to the substructure. Among clusters with no major substructure,
the typical fraction of galaxies in substructure is $\sim 10\%$ (e.g.,
Girardi et al.  \cite{gir97}; Guennou et al. \cite{gue14}).  In the
TOT sample, which extends out to $R\sim 2 R_{200}$ we formally
assigned  52\% of the galaxies to substructure.  However, we stress
that the precise number of objects assigned to the WNW clump is
largely uncertain owing to possible border effects.  The WNW
clump contains galaxies out to the western boundary of the sampled
region. We checked this point extending the size of our sample out to
$\sim 4 R_{200}$ with the addition of a random background field.  The
addition of a field equal to $5\%$ of the TOT sample does not change
the value of the relative density $\rho_{\rm S,WNW}$ of the WNW peak,
but reduces the number of the assigned WNW members.  The fraction of
WNW members in the TOT sample changes from WNW/TOT$\sim 23\%$ to
WNW/TOT$\sim 10\%$ leading to an important reduction in the estimated
fraction of galaxies in substructures.  In any case, our result agrees
with substructure being more important in the external cluster regions
(e.g., West \& Bothun \cite{wes90}).

\begin{figure*}
\centering
\resizebox{\hsize}{!}{\includegraphics{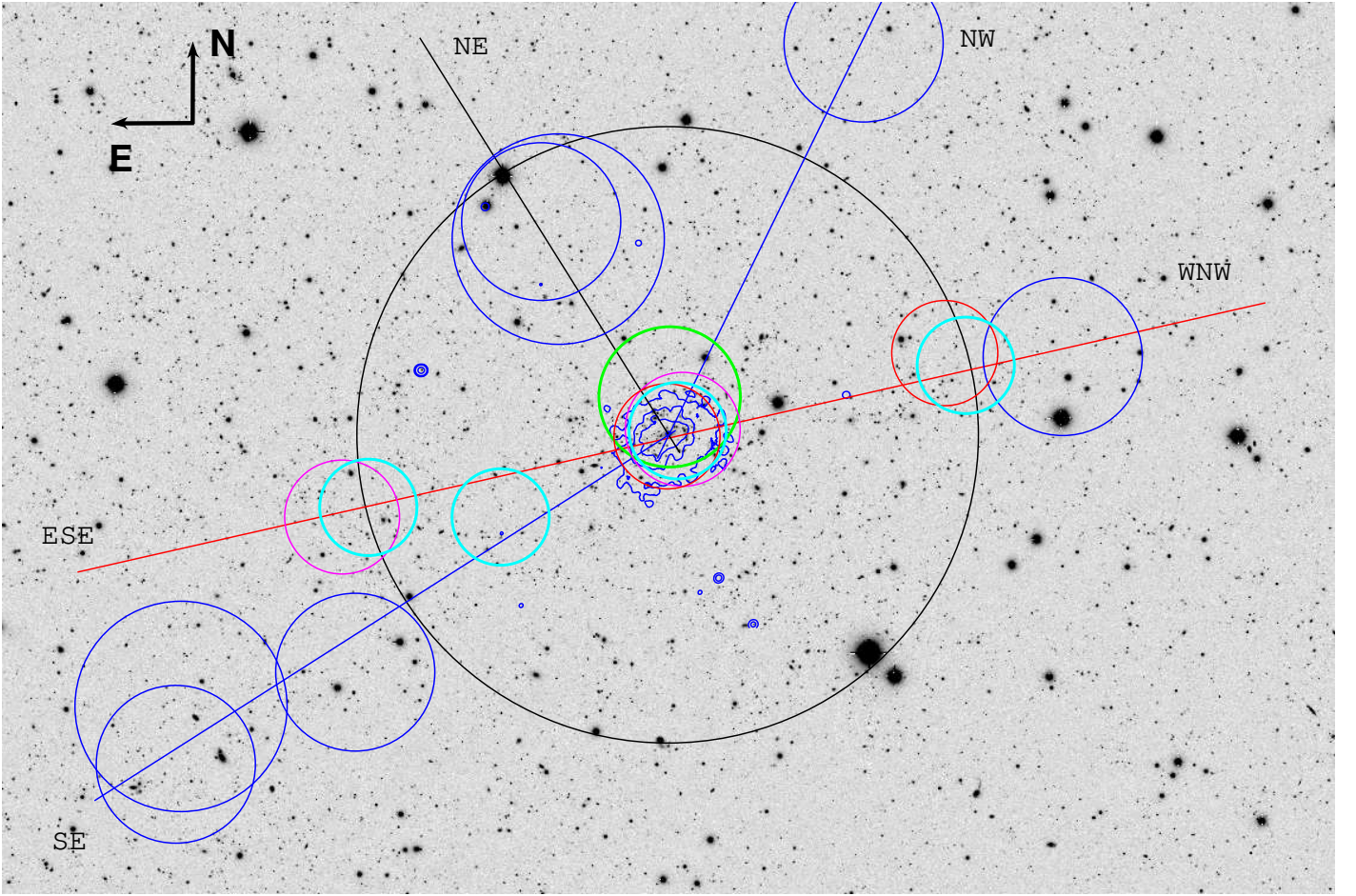}}
\caption{Subaru $R_C$ image of MACS1206 with superimposed features
  summarizing the results of the cluster structure analysis. Circles,
  in increasing size order, indicate the density peaks detected in the
  2D galaxy distribution of TOT (cyan), PAS (red), HDSr (magenta),
  wELG (green), ELG and vsELG (blue) samples (see
  Table~\ref{tabdedica2d} and Fig.~\ref{figk2zmul}); the NE and SEext
  clumps being also detected through our 3D analysis.  The lines
  indicate the likely directions of cluster accretion.  Blue contours
  are the smoothed contour levels of the X-ray surface brightness
  derived from Chandra archival data (in the 0.5-2 keV energy range;
  see also Ebeling et al. \cite{ebe09}).  The large black circle
  encloses the $R_{200}= 2$ \h $\sim 5.9$\arcmin\ region.  }
\label{figtot}
\end{figure*}

\subsection{Tracing a past phase of accretion}
\label{evol1}

Both the PAS and HDSr populations trace a cluster structure elongated
along the WNW-ESE direction suggesting that this is the direction of the
main accretion filament during the cluster formation.  Our
interpretation is that HDSr galaxies represent the galaxies more
affected by the last important phase of accretion, likely a past
merger or mergers along the WNW-ESE direction. In fact, the HDSr spectral
features reflect the evolution of galaxies having undergone starburst
or star-forming activity, which has been suppressed by some physical
process, possibly related to the high-density environment, and
observed 1-2 Gyr later (see, e.g., M04).  This time is comparable with
the crossing time of the present cluster ($R_{200}/\sigma_V\sim 1.8$
Gyr) indicating that the supposed merger(s) are in an
advanced phase. This also agrees with the fact that we do not see
relevant features of this past dynamical activity, since the
populations of the intervening subclusters have had time to mix. In
principle, we cannot exclude a phase of violent relaxation, since its
time scale is comparable to the dynamical time.  A time estimate of
1-2 Gyr is also consistent with the elongation of the X-ray isophotes
which is a long-lived phenomenon (e.g., Roettiger et
al. ~\cite{roe96}) and the asymmetry of the ICL, whose surviving time
can be comparable with one crossing time (Rudick et al. \cite{rud09}).

The observational evidence in favor of a past merger along the WNW-ESE
direction is the presence of a secondary dense peak in the
distribution of HDSr galaxies (ESE peak), confirmed in the
distribution of RedD galaxies, too (see Fig.~\ref{figk2zmul} and
Fig.~\ref{figk2zcol},  upper right panels).  The mean values of
EW(H$\delta$) of the RedU and RedD populations ($0.19\pm0.35$ \AA\ and
$1.10\pm0.24$ \AA) are only 2$\sigma$ different, but in the direction of
reinforcing our idea that the ESE peak is only traced by the  younger
population (RedD).  The HDSr ESE peak might be the remnant of a galaxy
system, now merged to form MACS1206.  However, it is formed only of
faint galaxies ($R_{\rm C}>$21.5 mag), lacking a dominant galaxy
which would have supported its identification with the remnant core of
an important subcluster.  Another point in favor of a past merger
along the WNW-ESE direction is the presence of a bright galaxy [at
  R.A.=$12^{\mathrm{h}}06^{\mathrm{m}}15\dotsec66$, Dec.=$-08\degree
  48\arcmm 21.8\arcs$ (J2000); also G2 in Grillo et al.
  \cite{gri14b}], which is the second brightest galaxy within 0.5 \h
from the cluster center, and which lies at the extremity of the asymmetric ICL
distribution, and seems to have strongly suffered from the interaction
with the BCG and/or the central cluster potential (see Fig.~6 of
Presotto et al. \cite{pre14}). This galaxy is also just at the border
of our inclusion in the HDSr sample [EW(H$_\delta$)=$3\pm1$ \AA].  A
third point is the WNW overdensity in the spatial distribution of
PASs, although the presence of a close overdensity of ELGs makes
 the interpretation difficult.

To understand the physical mechanisms generating HDSr galaxies, we note that,
at large scale, their spatial distribution follows that of PAS
galaxies. This agrees with the findings of Dressler et
al. (\cite{dre13}), who also suggested an evolutionary path of the
type PAS$\Rightarrow$PSB$\Rightarrow$PAS.  However, we cannot
completely agree with their scenario since our morphological analysis
based on S\'ersic indices shows that HDSr galaxies are intermediate
between PAS and ELG galaxies (see Fig.~\ref{figmatkwosersic}). The
intermediate morphology suggests that HDSr galaxies  are the
result of two mixed populations or that the physical process involved
in their transformation works on both their gas content and
morphology. The second interpretation favors a scenario of
merger or tidal interaction rather than a gas shocking or stripping. In
particular, the dense HDSr ESE concentration could be explained by a
mechanism similar to that suggested by Struck
(\cite{stru06}). According to this mechanism, the gravitational pull
of the cluster core on a small group falling through the core will
cause the group to shrink, increasing its density by an order of
magnitude, and as a result increasing the galaxy merger rate.

\subsection{Recent and ongoing galaxy infall}
\label{evol2}

Thanks to the large spatial coverage of the spectroscopic sample, the
ELG population can be studied in detail. We identify several clumps
(see Fig.~\ref{figk2zmul}, right lower panel). These galaxy clumps are
very loose and should be interpreted as overdensities in the LSS
filaments accreted to the cluster rather than gravitationally bound,
isolated groups.  These structures are generally external to $R_{200}$
with the exception of the NE clump which, being characterized
by a $\Delta V_{\rm NE,rf}\sim -550$ \kss, might be seen there because
of projection effects.  The ELG clumps we detect are likely the
cluster building blocks during the  ongoing galaxy infall.

N-body simulations by Balogh et al. (\cite{bal00}) and Moore et
al. (\cite{moo04}) follow the so-called backsplash or overshoot of
galaxies that have passed through the cluster and joined the
virialized systems: both these studies identify the zone between 1 and
2 $R_{200}$ as the overlap region where roughly half the galaxies are
members of the cluster and half are infalling. Since individual 
  ELG galaxy clumps are still detectable, we expect that their
members are likely infalling rather than backsplashing.  Considering
the region between 1 and 2 $R_{200}$, the ratio between galaxies that
belong to the ELG structures and the whole population is
$93/161\sim0.6$. Our result suggests that 60\% of galaxies have
  never passed through the cluster in rough agreement with the 50\%
  predicted by simulations by Balogh et al. (\cite{bal00}) and Moore
  et al. (\cite{moo04}).

When analyzing the geometry of the infall, the ELG clumps provide a few
preferred directions, traced in Fig.~\ref{figtot} roughly connecting
the clumps and the cluster center.  Figure~\ref{figtot} shows how both
the WNW-ESE and NW-SE directions are somewhat matched in the
elongation of X-ray isophotes, which are tilted going from the
internal to the external regions.  The net result of the secondary
infall is a somewhat rounder assembling than that due to the past
WNW-ESE accretion as suggested by the different ellipticity between
PAS and ELG galaxies (see Fig.~\ref{figell} and Table~\ref{tabell}).

\begin{figure}
\centering
\resizebox{\hsize}{!}{\includegraphics{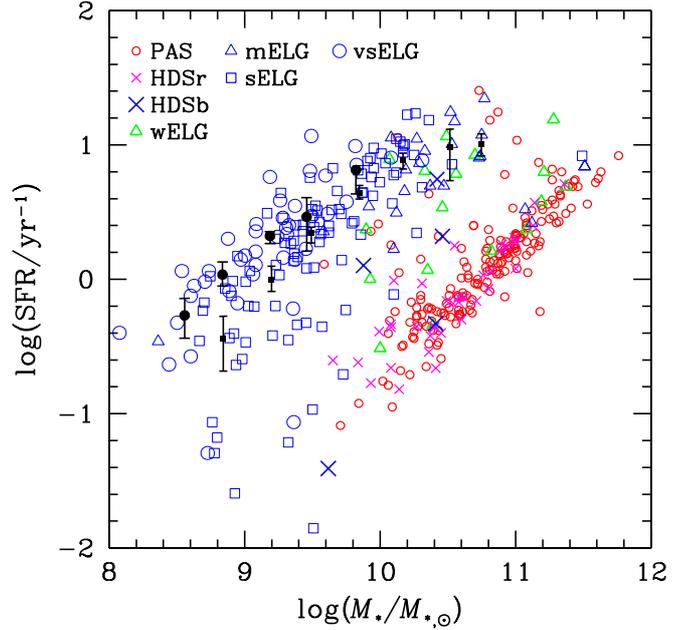}}
\caption{SFR vs stellar mass of cluster galaxies. Individual values
  are those computed by Annunziatella et al. (\cite{ann14}) and are
  shown here for different spectral classes. Black solid circles and small
  squares indicate biweight means with bootstrap errors for vsELGs and
  mELGs+sELGs considering three bins per order of magnitude in stellar
  mass.  }
\label{figSF}
\end{figure}

The galaxies of the vsELG class are spatially segregated with respect
to those of mELG+sELG classes, clustering in the NE and SEext
clumps. Figure~\ref{figSF} shows the values of stellar mass and SFR
that Annunziatella et al. (\cite{ann14}) computed for cluster member
galaxies. Comparing vsELGs and other ELGs in the $10^9\siml
M_*/M_{*,{\odot}} < 10^{10}$ range, we note that the specific SFR of
vsELGs is higher, particularly at low stellar masses. We interpret the
observed spatial segregation as the result of the same physical
process simultaneously acting in most galaxies within the parent
clump, which is enhancing SF in the NE and SEext clumps, or,
alternatively, quenching SF in other clumps.  In order to understand
whether this process is connected to the cluster or rather to a
pre-infalling environment, we analyzed the relative fraction of vsELGs
in the cluster and in the field.  To this end, we selected a
representative set of 776 field galaxies as non-member galaxies in the
$0.2<z<0.7$ redshift range, with spectral classification.  We compared
the fraction of vsELGs with respect to all ELGs.  We obtained that the
vsELG/ELG ratio in the cluster, 43/168 (26\%), is completely
consistent with the one in the field, 160/602 (27\%). This does
not support a scenario  where cluster environment is relevant in
  determining the vsELG/ELG ratio.

In the context of possible infalling galaxy structures, we stress that
our study focuses on galaxies belonging to the main cluster peak in
the redshift distribution, and does not consider foreground or
background redshifts peaks. In the study of Young et
al. (\cite{you14}), a few galaxies in a foreground redshift peak are
related to the presence of a significant interesting SZE feature, and
interpreted as a foreground or infalling group.

\subsection{Intermediate galaxy populations}
\label{evol3}

\begin{figure}
\centering
\includegraphics[width=9cm]{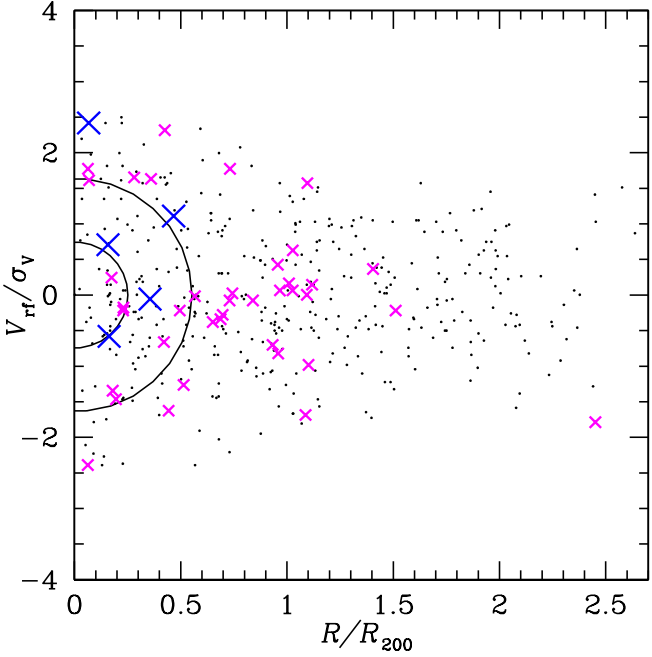}
\caption{Projected phase space as in Fig. \ref{figvdspec}, top panel,
  but with normalized units in position and velocity axes.  Small
  magenta crosses and large blue crosses indicate HDSr and HDSb
  galaxies.  Four of the five HDSb galaxies are  located within or
  close to the same ring-like region traced by the cluster PSBs
  analyzed by Muzzin et al. (\cite{muz14}); see text for details.}
\label{figmuzzin}
\end{figure}

Here we discuss the two intermediate populations, wELG and HDSb, which
according to our interpretation represent the fate of recently
in-fallen galaxies. The HDSb galaxies in our study are comparable to
the PSBs studied by Muzzin et al. (\cite{muz14}), which have very
strong EW(H$\delta$) [EW(H$\delta$)$>5$\AA\ for the average stacked
spectrum] and are defined with D$_n(4000)<1.45$, i.e., are blue PSBs
(see Fig. 2 of M04).  Figure~\ref{figmuzzin} shows that four of the five
HDSb galaxies in our sample are located within or close to the same
ring-like region in the projected phase space indicated by Muzzin et
al. (\cite{muz14}, see their Fig.~1). In that study PSBs are explained
with a rapid quenching of the order of $\siml 0.5$ Gyr after making
their first passage of $R\sim 0.5R_{200}$. The same time scale of
$\sim 0.5$ Gyr is invoked by M04 after the interruption of the SF in
the HDSb galaxies. In MACS1206, the five HDSb galaxies are elongated
towards the SE direction, approximately in the NW-SE direction of the
putative accretion flow (cf. Fig.~\ref{fig2dspec} and
Fig.~\ref{figtot}).  The wELG population appears distinct from the ELG
population, since wELGs populate inner cluster regions, and from the
PAS and HDSr populations since wELGs do not follow the WNW-ESE
elongation. Our interpretation is that wELGs are associated with a more
advanced phase in the ELG infalling process.

\section{Summary, conclusions, and perspectives}

To elucidate the scenario and physical mechanism leading to the
assembly of galaxy clusters, and specifically the transformation of
star-forming field galaxies to passive cluster galaxies, we present a
detailed spectrophotometric study of the massive cluster MACS1206 at
$z=0.44$ as part of the CLASH survey.  Our analysis is based on an
extensive spectroscopic dataset of 1920 galaxies over a field of ~640
arcmin$^2$ around MACS1206, which was the first observing campaign
conducted with VIMOS at the VLT as part of the ``CLASH-VLT'' ESO Large
Programme including 13 CLASH clusters in the southern sky. The
dataset is supplemented with multiband photometry based on Subaru
Suprime-Cam high-quality imaging.  We combine galaxy velocities and
positions to select 445 cluster galaxies based on kinematic
information alone.  Using spectral absorption and emission features,
as well as galaxy colors, we classify 412 galaxies in seven spectral classes
ranging from purely passive galaxies to those with strong H$\delta$
absorption (red and blue), and emission line galaxies (four classes
from weak to very strong lines). We then analyze the cluster structure
and distribution of these different galaxy populations in projected
phase-space out to a radius $\sim 2R_{200}$ (4 \hh). Our specific
results can be summarized as follows:

\begin{itemize}

\item MACS1206 appears as a large-scale relaxed system, particularly
  within $R_{200}$, with a few, low overdensities detected in the
  projected galaxy distribution (relative density $\siml 20\%$). The
  whole galaxy population does not show velocity gradients or
  evidence of significant deviations in local mean velocities. The
  main large-scale feature is the elongation along the WNW-ESE
  direction (PA $\sim$117\degreee).

\item Passive galaxies and red, strong H$\delta$ galaxies trace the cluster
  center and the WNW-ESE elongated structure. The latter also cluster
  around a secondary, very dense peak $2$ \h at ESE.

\item The emission line galaxies are mostly located in several loose
  structures, mostly outside $R_{200}$. Two of these structures are
  also detected in the 3D projected phase space having LOS rest-frame
  velocities $\Delta V_{\rm rf,NE}\sim -550$ \kss and $\Delta V_{\rm
    rf,SEext}\sim +550$ \kss.

\item The clumps of emission line galaxies trace several directions,
  with the net result of a more symmetrical distribution than that
  traced by early-type galaxies (the difference in ellipticity is a
  factor of 1.5-2).

\end{itemize}

By studying the spatial and velocity distributions of galaxy
populations of different spectral types, we can piece together a likely
scenario for the galaxy assembly history of this massive system. The
oldest and main component of MACS1206, made of early-type galaxies,
clearly traces the main phase of cluster formation. However, a more
recent anisotropic accretion phase that occurred 1-2 Gyr before the epoch of
observation is apparent in the distribution of red strong H$\delta$
galaxies along the prominent WNW-ESE filament, including a strong
overdensity of these evolved post-SF galaxies (ESE peak), possibly the remnant
of an accreted group. In addition, we identified a secondary prominent young
component in the external regions of MACS1206 populated by late-type
galaxies, which trace the ongoing infall.

Our study shows that large spectroscopic samples of cluster galaxies
extending to $\simg 2 R_{200}$ allow us to link the cluster
substructure with distinct galaxy populations, whose spectral types
can be used to age-date their underlying stellar populations and SF
history. This ultimately allows us to reconstruct the evolution of
galaxies and their parent cluster, thus going beyond the simplistic
concept of radial accretion and following cluster formation through
their building blocks.  We plan to extend this analysis to other
clusters in the CLASH-VLT survey, which include systems of different
dynamical states and over an interesting redshift range. A more
refined spectral classification will also be possible in clusters in
this program, which include $\sim\! 7000$ spectroscopic members.

\begin{acknowledgements}
We thank the referee for invaluable comments.  This research is partly
supported by the MIUR PRIN2010-2011 (J91J12000450001) and PRIN INAF
2014.  I.B. acknowledges funding from the European Union Seventh
Framework Programme (FP7/2007-2013) under grant agreement n.~267251
``Astronomy Fellowships in Italy'' (astroFit).  The Dark Cosmology
Centre is funded by the DNRF.  B.S. and V.P. acknowledge a grant from ``Consorzio per la Fisica - Trieste”.  R.D. gratefully acknowledges the
support provided by the BASAL Center for Astrophysics and Associated
Technologies (CATA), and by FONDECYT grant N. 1130528.
A.F. acknowledges support by INAF through VIPERS grants PRIN 2008 and
PRIN 2010.  Support for A.Z. was provided by NASA through Hubble
Fellowship grant \#HST-HF2-51334.001-A awarded by STScI.  The research is based in
part on data collected at Subaru Telescope, which is operated by the
National Astronomical Observatory of Japan, and obtained from the
SMOKA, which is operated by the Astronomy Data Center, National
Astronomical Observatory of Japan.

\end{acknowledgements}


\begin{thebibliography}{}

\bibitem[1978]{aar78} Aarseth, S. J., \& Binney, J. 1978, \mnras, 185, 227

\bibitem[1996]{abr96} Abraham, R. G., Smecker-Hane, T. A., Hutchings, J. B. et al. 1996, \apj, 471, 694

\bibitem[1998]{ada98} Adami, C., Biviano, A., \& Mazure, A. 1998, \aap, 331, 439

\bibitem[2014]{ann14} Annunziatella, M. Biviano, A., Mercurio, A., et al.  2014, A\&A, 571, A80

\bibitem[1994]{ash94} Ashman, K. M., Bird, C. M., \& Zepf, S. E. 1994, \aj, 108, 2348

\bibitem[2002]{ath02} Athreya, R. M., Mellier, Y., van Waerbeke, L.,
et al. 2002, \aap, 384, 743

\bibitem[1977]{bai77} Baier, F. W., \& Ziener, R. 1977, Astron. Nachr., 298, 87

\bibitem[1999]{bal99} Balogh, M. L., Morris, S.L., Yee, H.K.C., Carlberg, R.G., \& Ellingson,
E. 1999, \apj, 527, 54

\bibitem[2000]{bal00} Balogh, M. L., Navarro, J. F., \&  Morris, S. L. 2000, \apj,  540, 113 

\bibitem[2012]{bar12} Barden, M.,  H\"aussler, B. Peng, C. Y., McIntosh, D. H., \& Guo, Y. 2012, \mnras, 422, 449

\bibitem[2005]{bar05} Barden, M., Rix, H.-W., Somerville, R. S. et al. 2005, \apj, 635, 959


\bibitem[1996]{bar96} Barger, A., Aragon-Salamanca, A., Ellis, R., et al. 1996, MNRAS, 279, 1

\bibitem[2011]{bar11} Barrena, R., Girardi, M., Boschin, W., et al. 2011, \aap, 529, A128


\bibitem[2006]{bas06} Basilakos, S., Plionis, M., Yepes, G., Gottlöber, S., \&
Turchaninov, V. 2006, \mnras, 365, 539

\bibitem[1990]{bee90} Beers, T. C., Flynn, K., \& Gebhardt, K. 1990, \aj, 100, 32

\bibitem[1999]{bek99} Bekki, K. 1999, \apj, 510, L15


\bibitem[2010]{bek10} Bekki, K., Owers, M. S., \& Couch, W. J. 2010, \apjl, 718, L27


\bibitem[1993]{bir93} Bird, C. M., \& Beers, T. C. 1993, \aj, 105, 1596


\bibitem[1996]{biv96} Biviano, A., Durret, F., Gerbal, D. et al. 1996, \aap, 311, 95

\bibitem[1997]{biv97} Biviano, A., Katgert, P., Mazure, A. et al. 1997, \aap, 321, 84

\bibitem[2002]{biv02} Biviano, A., Katgert, P., Thomas, T., \& Adami, C. 2002, \aap, 387, 8

\bibitem[1992]{biv92} Biviano, A., Girardi, M., Giuricin, G., Mardirossian, F., \& Mezzetti, M. 1992, \apj, 396, 35

\bibitem[2013]{biv13} Biviano, A., Rosati, P., Balestra, I., et al.
2013, \aap, 558, A1 (B13)

\bibitem[2004]{boe04} B\"ohringer, H., Pratt,  Schuecker, P.; Guzzo, L., et al. 2004, 425, 367

\bibitem[2010]{boe10} B\"ohringer, H., Pratt, G. W., Arnaud, M., et al. 2010, \aap, 514 A32

\bibitem[1983]{bru83} Bruzual, G. 1983, \apj, 273, 105

\bibitem[2002]{buo02} Buote, D. A. 2002, in ``Merging Processes in Galaxy Clusters'', eds. L. Feretti, I. M. Gioia, \& G. Giovannini (The Netherlands, Kluwer Ac. Pub.): Optical Analysis of Cluster Mergers

\bibitem[1984]{but84} Butcher, H., \& Oemler, A., Jr. 1984, \apj, 285, 426

\bibitem[1997]{cal97} Caldwell, N., \& Rose, J. A. 1997, \aj, 113, 492

\bibitem[1980]{car80} Carter, D., \& Metcalfe, J. 1980, \mnras, 191, 325

\bibitem[2014]{coh14} Cohen, S. A., Hickox, R. C., Wegner, G. A., Einasto, M., \&  Vennik, J. 2014, \apj, 783, 136

\bibitem[1999]{col99} Colberg, J. M., White, S. D. M., Jenkins, A., \& Pearce, F. R. 1999, 
MNRAS, 308, 593

\bibitem[1998]{col98} Colberg, J. M., White, S. D. M., Macfarland, T. J. et al. 1998, in  
{\it Wide Field Surveys in Cosmology} (Edition Frontiers), p. 247

\bibitem[1996]{col96} Colless, M., \& Dunn, A. 1996, \apj, 458, 435

\bibitem[1987]{cou87} Couch, W. J., \& Sharples, R. M. 1987, MNRAS, 229, 423

\bibitem[2002]{czo02} Czoske, O., Moore, B., Kneib, J.-P., \& Soucail, G. 2002, \aap, 386, 31

\bibitem[2002]{dah02} Dahle, H., Kaiser, N., Irgens, R. J., Lilje, P. B., \& Maddox, S. J. 2002, \apjs, 139, 313

\bibitem[1980]{dan80} Danese, L., De Zotti, C., \& di Tullio, G. 1980, \aap, 82, 322

\bibitem[2007]{dem07} Demarco, R., Rosati, P., Lidman, C., et al. 2007, \apj, 663, 164

\bibitem[1996]{den96} den Hartog, R., \& Katgert, P. 1996, \mnras, 279, 349

\bibitem[2012]{die12} Dietrich, J. P., Werner, N., Clowe, D., et al. 2012 \nat, 487, 202

\bibitem[1998] {don98} Donahue, M., Voit, G. M., Gioia, I. M., et al. 1998, \apj, 502, 550

\bibitem[1980]{dre80} Dressler, A. 1980, \apj, 236, 351

\bibitem[1983]{dre83} Dressler, A. \& Gunn, J. E. 1983, \apj, 270, 7

\bibitem[2013]{dre13} Dressler, A., Oemler, A. Jr., Poggianti, B. M., et al.
2013, \apj, 770, 62

\bibitem[2009]{dre09} Dressler, A., Rigby, J, Oemler, A. Jr., et al. 2009, \apj, 693, 140

\bibitem[1988]{dre88} Dressler, A., \& Shectman, S. A. 1988, \aj, 95, 985

\bibitem[1997]{dre97} Dressler, A., Oemler, A. Jr., Couch, W. J., et al. 1997, \apj, 490, 577
 
\bibitem[1999]{dre99} Dressler, A., Smail, I., Poogianti, B. M., et al.
1999, \apj, 122, 51

\bibitem[2009]{ebe09} Ebeling, H., Ma, C. J., Kneib, J.-P. Kneib, et al. 2009,
\mnras, 395, 1213

\bibitem[2008]{fad08} Fadda, D., Biviano, A.,
Marleau, F. R., Storrie-Lombardi, L. J., \& Durret, F. 2008, \apj, 672, L9

\bibitem[1996]{fad96} Fadda, D., Girardi, M., Giuricin, G., Mardirossian, F., \& Mezzetti, M. 1996, \apj, 473, 670

\bibitem[1987]{fas87} Fasano, G., \& Franceschini, A. 1987, \mnras, 225, 155

\bibitem[2011]{fas11} Fassbender, R., B\"ohringer, H., Nastasi, A., et al. 2011, New Journal of Physics, 13, 125014

\bibitem[2002]{fer02} Feretti, L., Gioia I. M., and Giovannini G. eds., 2002b, Astrophysics and Space Science Library, vol. 272, ``Merging Processes in Galaxy Clusters'', Kluwer Academic Publisher, The Netherlands

\bibitem[2005]{fer05} Ferrari, C., Benoist, C., Maurogordato, S., Cappi, A., \& Slezak, E. 2005, \aap, 430, 19   

\bibitem[2008]{fis08} Fisher, D. B., \& Drory, N.  2008, \aj, 136, 773

\bibitem[2005]{fri05} Fritz, A., Ziegler, B. L., Bower, R. G., Smail, I., \& Davies, R. L. 2005, \mnras, 358, 233

\bibitem[1999]{fuj99} Fujita, Y., Takizawa, M., Nagashima, M., \& Enoki, M., \pasj, 51, L1

\bibitem[1991]{geb91} Gebhardt, K., \& Beers, T. C. 1991, \apj, 383, 72

\bibitem[1982]{gel82} Geller, M. J., \& Beers, T. C. 1982, \pasp, 94, 421

\bibitem[2004]{ger04} Gerken, B., Ziegler, B., Balogh, M., et al. 2004, \aap,
421, 59

\bibitem[2005]{gil05} Gill, S. P. D., Knebe, A., \& Gibson , B. K. 2005, 
\mnras, 356, 1327

\bibitem[1999] {gio99} Gioia, I. M.,  Henry, J. P., Mullis, C. R.,
Ebeling, H.; Wolter, A. 1999, \aj, 117, 2608

\bibitem[2011]{gir11} Girardi, M., Bardelli, S., Barrena, R., et al. 2011, \aap, 536, A89

\bibitem[2002]{gir02} Girardi, M., \& Biviano, A. 2002, in ``Merging Processes in Galaxy Clusters'', eds. L. Feretti, I. M. Gioia, \& G. Giovannini (The Netherlands, Kluwer Ac. Pub.): Optical Analysis of Cluster Mergers


\bibitem[2006]{gir06} Girardi, M., Boschin, W., \& Barrena, R. 2006, \aap, 455, 45

\bibitem[1997]{gir97} Girardi, M., Escalera, E., Fadda, D., et al. 1997, \apj, 482, 11

\bibitem[1996]{gir96} Girardi, M., Fadda, D., Giuricin, G. et al. 1996, \apj, 457, 61

\bibitem[1998]{gir98} Girardi, M., Giuricin, G., Mardirossian, F., Mezzetti, M., \& Boschin, W. 1998, \apj, 505, 74

\bibitem[2003]{gom03} G\'omez, P. L., Nichol, R. C., Miller, C. J.
et al. 2003, \apj, 584, 210

\bibitem[1994]{gon94} Gonz\'alez-Casado, G., Mamon, G. A., \& Salvador-Sol\'e, E. 1994, \apjl, 433, 61

\bibitem[2014b]{gri14b} Grillo, C.,  Gobat, R., Presotto, V., et al. 2014,
\apj, 786, 11

\bibitem[2014a]{gri14a} Grillo, C.,  Suyu, S. H., Rosati, P., et al. 2014,
\apj, 800, 38

\bibitem[2014]{gue14} Guennou, L., Adami, C., Durret, F., et al. 2014, \aap, 516, A112 

\bibitem[1961]{gue61} Guest, P. G. 1961, {\em Numerical Methods
    of Curve Fitting} (Cambridge: Cambridge Univ. Press).


\bibitem[1997]{ham97} Hammer, F., Flores, H., Lilly, S. J., et al. 1997, \apj, 481, 49

\bibitem[1997]{ish97} Ishizaka, C. 1997, \apss, 254, 233

\bibitem[2005]{jel05} Jeltema, T. E., Canizares, C. R., Bautz, M. W., \& Buote, D. A. 2005, \apj, 624, 606

\bibitem[1999]{jon99} Jones, C., \& Forman, W. 1999, \apj, 511, 65

\bibitem[2004]{jon04} Jones, D. H., Saunders, W., Colless, M., et al. 2004, MNRAS, 355, 747

\bibitem[2006]{lam06} Lamareille, F., Contini, T., Le Borgne, J.-F., et al. 2006, \aap, 448, 893

\bibitem[1982]{led82} Ledermann, W. 1982, Handbook of Applicable Mathematics (New York: Wiley),
Vol.6.

\bibitem[2013]{lem13} Lemze, D., Postman, M., Genel, S., et al. 2013, \apj, 776, 91

\bibitem[2002]{lew02} Lewis, I., Balogh, M., De Propris, R.  et al. 2002, \mnras, 334, 673

\bibitem[2010]{ma10} Ma, C.-J., Ebeling, E., Marshall, P., \& Schrabback, T. 2010, \mnras,
406, 121

\bibitem[2011]{mah11} Mahajan, S., Mamon, G. A., \ Raychaudhury, S. 2011, \mnras, 416, 2882

\bibitem[2012]{mah12} Mahajan, S., Raychaudhury, S., \& Pimbblet, K. A. 2012, \mnras, 427, 1252

\bibitem[2011]{mau11} Maurogordato, S., Sauvageot, J. L., Bourdin, H., et al. 2011, \aap, 525, A79

\bibitem[1977]{mel77} Melnick, J., \& Sargent, W. L. 1977, \apj, 215, 401

\bibitem[2004]{mer04} Mercurio, A., Busarello, G., Merluzzi, P., et al.  2004, \aap, 424, 79 (M04)

\bibitem[2008]{mer08} Mercurio, A., La Barbera, F., Haines, C. P., et al.  2008, MNRAS, 387, 1374 

\bibitem[2004]{moo04} Moore, B., Diemand, J., \& Stadel, J. 2004 in Outskirts of Galaxy Clusters: Intense Life in the Suburbs, ed. A. Diaferio, IAU Colloquium 
\#195, p.513

\bibitem[1977]{mos77} Moss, C. \& Dickens, R. J. 1977, 178, 701 

\bibitem[2014]{mun14} Munari, E., Biviano, A., \& Mamon, G. 2014, \aap,
566, 68

\bibitem[2004]{mus97} Mushotzky, R. F., \& Scharf, C. A. 1997, \apjl, 482, L13

\bibitem[2014]{muz14} Muzzin, A., van der Burg, R. F. J., McGee, S. L. et al. 2014, \apj, 796 65

\bibitem[2009]{non09} Nonino, M., Dickinson, M., Rosati, P., et al. 2009, \apjs, 183, 244

\bibitem[2009]{oem09} Oemler, A. Jr., Dressler, A., Kelson, D.,
et al. 2009, \apj, 693, 152

\bibitem[2005]{owe05} Owen, M. S., Ledlow, M. J., Keel, W. C., Wang, Q. D., \& Morrison,  G. E. 2005, \aj, 129, 31

\bibitem[2011]{owe11} Owers, M. S., Randall, S. W., Nulsen, P. E., et al. 2011, \apj, 728, 27

\bibitem[1996]{pin96} Pinkney, J., Roettiger, K., Burns, J. O., \& Bird, C. M. 1996, \apjs, 104, 1

\bibitem[1993]{pis93} Pisani, A. 1993, \mnras, 265, 706

\bibitem[1996]{pis96} Pisani, A. 1996, \mnras, 278, 697

\bibitem[2002]{pli02} Plionis, M.,  \& Basilakos, S.    2002, \mnras, 329, L47

\bibitem[1996]{pog96} Poggianti, B. M., \& Barbaro, G. 1996, A\&A, 314, 379

\bibitem[1997]{pog97} Poggianti, B. M., \& Barbaro, G. 1997, A\&A, 325, 1025

\bibitem[2004]{pog04} Poggianti, B. M., Bridges, T. J., Komiyama, Y., et al.
 2004, \apj, 601, 197

\bibitem[1999]{pog99} Poggianti, B. M., Smail, I., Dressler, A., et al. 1999, \apj, 518, 576
 
\bibitem[2000]{pog00} Poggianti, B. M., \& Wu, H. 2000, \apj, 529, 157


\bibitem[2009]{pog09} Poggianti, B. M., Arag\'on-Salamanca, A., Zaritsky, D.,
et al. 2009, \apj, 693, 112

\bibitem[2007]{por07} Porter, S. C., \& Raychaudhury, S. 2007, \mnras, 375, 1409

\bibitem[2012]{pos12} Postman, M., Coe, M., Ben\'itez, N., et al. 2012, \apjs, 199, 25

\bibitem[2014]{pre14} Presotto, V., Girardi, M., Nonino, M., et al. 2014, \aap, 565, A126

\bibitem[1992]{pre92} Press, W. H., Teukolsky, S. A., Vetterling, W. T., \& Flannery, B. P. 1992, in Numerical Recipes (Second Edition), (Cambridge University Press)


\bibitem[2007]{ram07} Ramella, M., Biviano, A., Pisani, A., et al. 2007, \aap, 470, 39

\bibitem[2012]{ras12} Rasmussen, J, Mulchaey, J. S., Bai, L. et al. 2012, \apj, 757, 122

\bibitem[2013]{rin13} Rines, K., Geller, M. J., Diaferio, A., \& Kurtz, M. J.
2013, \apj, 767, 15     

\bibitem[2005]{rin05} Rines, K., Geller, M. J., Kurtz, M. J., \& 
Diaferio, A., \& Kurtz, M. J. 2005, \aj, 130, 1482      

\bibitem[1996]{roe96} Roettiger, K., Burns, J. O., \& Loken, C. 1996, \apj, 473, 651

\bibitem[2014]{ros14} Rosati, P., Balestra, I., Grillo, C. et al.  2014, The Messenger,  158, 48

\bibitem[2009]{rud09} Rudick, C. S., Mihos, J. C., Frey, L. H., \& McBride, C. K. 2009, \apj, 699, 1518

\bibitem[1986]{sar86} Sarazin, C. L. 1986, {\em Reviews of Modern Physics}, 58, 1

\bibitem[1998]{sch98} Schlegel, D. J., Finkbeiner, D. P., \& Davis, M. 1998, \apj, 500, 525

\bibitem[2005]{sco05} Scodeggio, M., Franzetti, P., Garilli, B., et al. 2005, PASP, 117, 1284

\bibitem[2005]{smi05} Smith, G. P.,  Kneib, J.-P., Smail, I., et al. 2005, \mnras , 359, 417

\bibitem[1989]{sod89} Sodr\'e, L. Jr., Capelato, H. V., Steiner, J. E., Mazure, A. 1989, \aj, 97, 1279

\bibitem[2006]{stru06} Struck, C. 2006, in ``Atrophysics Update 2'', ed. J. W. Mason (Heidelberg: Springer), 115 

\bibitem[1972]{tam72} Tammann, G. A. 1972, \aap, 21, 355

\bibitem[1998]{tra98} Trager, S.C., Worthey, G., Faber, S.M., Burstein D., \& Gonzalez J.J., 1998, \apjs, 116, 1

\bibitem[2003]{tre03} Treu, T., Ellis, R. S., Kneib, J.-P. et al. 2003, \apj, 591, 53

\bibitem[2012]{ume12} Umetsu, K., Medezinski, E.,  Nonino, M. et al. 2012, \apj, 755, 56 

\bibitem[2013]{vin13} Vijayaraghavan, R., \& Ricker, P. M. 2013, \mnras, 435, 2713 

\bibitem[2013]{wen13} Wen, Z. L., \& Han, J. L. 2013, \mnras, 436, 275

\bibitem[1990]{wes90} West, M. J., \& Bothun , G. D. 1990, \apj, 350, 36

\bibitem[1996]{whi96} White, S. D. 1996 in {\em Gravitational dynamics}, Proceedings of the 36th Herstmonceux conference, in honour of Professor D. Lynden-Bell's 60th birthday, Cambridge, UK, August 7-11, 1995, Cambridge Univeristy Press, 
eds. Lahav, O., Terlevich, E., Terlevich, R. J., p.121

\bibitem[1993]{whi93} Whitmore, B. C., Gilmore, D. M., \& Jones, C. 1993, \apj, 407, 489
 
\bibitem[1994]{wor94} Worthey, G., Faber, S.M., Gonzalez, J.J., \& Burstein, D. 1994, \apj, 94, 687

\bibitem[1997]{wor97} Worthey, G., \& Ottaviani, D.L. 1997, \apjs, 111, 377

\bibitem[2014]{you14} Young, A. H., Mroczkowski, T.,  Romero, C. 2014, \apj, subm., preprint arXiv:1411.0317

\bibitem[1993]{zab93} Zabludoff, A. I., \& Franx, M. 1993, \aj, 106, 1314

\bibitem[2009]{zha09} Zhang, Y.-Y., Reiprich, T. H., Finoguenov, A., Hudson, D. S., \& Sarazin, C. L. 2009, \apj, 699, 1178

\bibitem[2012]{zit12} Zitrin, A., Rosati, P., Nonino, M., et al. 2012, \apj, 749, 97

\end{thebibliography}
\end{document}